\documentclass[sn-mathphys,Numbered]{sn-jnl}

\usepackage{graphicx}%
\usepackage{multirow}%
\usepackage{amsmath,amssymb,amsfonts}%
\usepackage{amsthm}%
\usepackage{mathrsfs}%
\usepackage[title]{appendix}%
\usepackage{xcolor}%
\usepackage{textcomp}%
\usepackage{manyfoot}%
\usepackage{booktabs}%
\usepackage{algorithm}%
\usepackage{algorithmicx}%
\usepackage{algpseudocode}%
\usepackage{listings}%
\usepackage{tikz}%
\usepackage{subcaption}%
\usepackage{siunitx}

\theoremstyle{thmstyleone}%
\theoremstyle{thmstyletwo}%

\theoremstyle{thmstylethree}%

\begin{document}


\title[An automated algorithmic method to mitigate long-term variations in the efficiency of the GRAPES-3 muon telescope]
      {An automated algorithmic method to mitigate long-term variations in the efficiency of the GRAPES-3 muon telescope}


\author[1]{\fnm{S.} \sur{Paul}}
\author[2]{\fnm{K.P.} \sur{Arunbabu}}
\author[1]{\fnm{M.} \sur{Chakraborty}}
\author[1]{\fnm{S.K.} \sur{Gupta}}
\author*[1]{\fnm{B.} \sur{Hariharan}}\email{89hariharan@gmail.com}
\author[3]{\fnm{Y.} \sur{Hayashi}}
\author[1]{\fnm{P.} \sur{Jagadeesan}}
\author[1]{\fnm{A.} \sur{Jain}}
\author[4]{\fnm{P.} \sur{Jain}}
\author[1]{\fnm{M.} \sur{Karthik}}
\author[3]{\fnm{S.} \sur{Kawakami}}
\author[5]{\fnm{H.} \sur{Kojima}}
\author[1]{\fnm{K.} \sur{Manjunath}}
\author[1]{\fnm{P.K.} \sur{Mohanty}}
\author[1]{\fnm{S.D.} \sur{Morris}}
\author[6]{\fnm{Y.} \sur{Muraki}}
\author[1]{\fnm{P.K.} \sur{Nayak}}
\author[7]{\fnm{T.} \sur{Nonaka}}
\author[5]{\fnm{A.} \sur{Oshima}}
\author[1]{\fnm{D.} \sur{Pattanaik}}
\author[1]{\fnm{B.} \sur{Rajesh}}
\author[1]{\fnm{M.} \sur{Rameez}}
\author[1]{\fnm{K.} \sur{Ramesh}}
\author[1]{\fnm{B.S.} \sur{Rao}}
\author[1]{\fnm{L.V.} \sur{Reddy}}
\author[5]{\fnm{S.} \sur{Shibata}}
\author[8]{\fnm{K.} \sur{Tanaka}}
\author[4]{\fnm{F.} \sur{Varsi}}
\author[1]{\fnm{M.} \sur{Zuberi}}

\affil[1]{\orgname{Tata Institute of Fundamental Research}, \orgaddress{\street{Homi Bhabha Road}, \city{Mumbai 400005}, \country{India}}}
\affil[2]{\orgname{Cochin University of Science and Technology}, \orgaddress{\city{Kochi 682022}, \country{India}}}
\affil[3]{\orgname{Graduate School of Science}, \orgaddress{\street{Osaka City University}, \city{Osaka 558-8585}, \country{Japan}}}
\affil[4]{\orgname{Indian Institute of Technology Kanpur}, \orgaddress{\city{Kanpur 208016}, \country{India}}}
\affil[5]{\orgname{College of Engineering}, \orgaddress{\street{Chubu University}, \city{Kasugai, Aichi 487-8501}, \country{Japan}}}
\affil[6]{\orgname{Institute for Space-Earth Environmental Research}, \orgaddress{\street{Nagoya University}, \city{Nagoya 464-8601}, \country{Japan}}}
\affil[7]{\orgname{Institute for Cosmic Ray Research}, \orgaddress{\street{University of Tokyo}, \city{Chiba 277-0882}, \country{Japan}}}
\affil[8]{\orgname{Graduate School of Information Sciences}, \orgaddress{\street{Hiroshima City University}, \city{Hiroshima 731-3194}, \country{Japan}}}


\abstract{The GRAPES-3 large area muon telescope with its sixteen independent modules records the high energy ($>$1\,GeV) muons continuously over 2.3\,sr of the sky. However, the recorded muon rates are contaminated by instrumental effects and instabilities spanning both short- and long-timescales, such as variations in the efficiency of the detector. We present an automated, algorithmic method, which employs Bayesian blocks to discretize the data stream into periods and exploits the correlations among the sixteen independent modules of the muon telescope to separate the impact of these instrumental problems from those originating in physical effects of interest, allowing the Savitzky-Golay filter to be employed to mitigate the former. Compared to legacy methods, this method is less dependent on subjective input from experimental operators and provides a data stream free of all known instrumental effects over calendar years. The muon rate obtained with the new method shows a fairly better correlation with neutron monitor data, than that obtained with the legacy method.}

\maketitle


\section{Introduction \label{intro}}

The \textbf{G}amma \textbf{R}ay \textbf{A}stronomy at \textbf{P}eV \textbf{E}nergie\textbf{S} - phase \textbf{3} (GRAPES-3) is a ground-based cosmic ray (CR) experiment designed primarily to record extensive air showers (EASs) in the energy (of the primary) range of 10\,TeV--10\,PeV and more. Its large area (560\,m$^2$) muon telescope, hereafter referred to as G3MT (GRAPES-3 muon telescope) samples the muon content in the EASs, enabling the gamma-initiated EASs to be differentiated from hadron-initiated ones. In addition, it also records the angular muon rate above a GeV, produced by EASs initiated by primaries with energies of 10 GeV and above, enabling the study of near-Earth and atmospheric phenomena as well as the modulation of CRs by the solar system. In continuous operation since the year 2000, data produced by G3MT has enabled path-breaking studies of the long-term modulation of CR flux \cite{Arunbabu_2017,Kojima_2024}, atmospheric acceleration \cite{Hari_2019}, interplanetary coronal mass ejections \cite{Mohanty_2016_2,Mohanty_2018}, and solar and heliospheric phenomena \cite{Arunbabu_2009,Arunbabu_2013,Arunbabu_2015}, as well as measurements of the spectrum and anisotropy in the arrival directions of CRs~\cite{Varsi_2024,Chakraborty_2024}. These studies allow probing of the origin and propagation of CRs within the atmosphere, solar system, and Milky Way, respectively.

However, before any physics studies can be performed, it is necessary to mitigate the variations in G3MT data originating from known systematic and instrumental effects. Atmospheric pressure is one of the major sources of periodic systematic variations, and can be easily corrected for if precise measurements of the atmospheric pressure on site are available, whereas correcting for instrumental effects is nontrivial since it is challenging to confidently decouple such efforts from physical phenomena. The instrumental effects can be broadly divided into (i) short- and (ii) long-term effects, typically attributed to momentary and localized issues at various stages of signal processing, and slow decline in detectors' efficiency, respectively. The former introduces incurable errors in the data from individual modules of G3MT in most scenarios, with the time span ranging anywhere from a few minutes to several hours. The latter introduces a slow decline in the overall performance of specific G3MT modules and can be observed on longer time scales (i.e., ranging from a few months to years) -- requiring corrective measures before the data can be used for studies of long-term CR modulation \cite{Mohanty:2017umr}. 

Traditionally, this has been achieved by identifying a reference module based on the judgment of the experiment operators~\cite{Mohanty:2017umr}. While this has so far served the purpose, as the experiment continues to operate into the future, automated, algorithmic methods that are less susceptible to human subjectivity become necessary. Since G3MT is a collection of sixteen independent telescope modules, the correlation between the muon rates in the different modules can be used as a reference to overcome the above-mentioned issues without affecting the variations caused by physical phenomena. To address this, methods have been developed to model both short- and long-term instrumental variations, with corresponding corrections implemented. This report provides a detailed summary of these efforts. The next section discusses the GRAPES-3 experiment and its detectors. Sec.\,\ref{causes} discusses various types of systematic and instrumental effects and their respective causes. The latter Sec.\,\ref{method} discusses the streamlined and automated approach to parameterization and correction of systematic and instrumental effects. Once these effects are corrected, the independent muon module rates may be combined to form omni and multi-directional muon rate profiles for physics studies. These methods are suitable for application to data from other experiments that record CR flux, and has the potential to reduce systematic uncertainties.


\section{The GRAPES-3 muon telescope \label{g3mt}}

GRAPES-3 is a ground-based observatory located in the Nilgiri mountains in Ooty, India (11.4$^\circ$N, 76.7$^\circ$E, 2200\,m above mean sea level). It consists of two detector elements, (i) an array of 400 plastic scintillator detectors (G3SD) spread over, 25000\,m$^2$ \cite{Gupta_2005} and (ii) a large area (560\,$^2$) tracking muon telescope (G3MT) \cite{Hayashi_2005}. The G3SD is tuned to record EASs produced by the CRs in a broad energy range of 10\,TeV--10\,PeV, bridging the reach of direct and indirect detection experiments~\cite{Hari_2020}. The G3MT consists of sixteen independent muon modules (35\,$^2$ each) arranged in four logical groups called muon super-modules (Fig\,2, 3, and 4 of \cite{Hayashi_2005}). Each muon module is built using four layers of proportional counters (PRCs). The PRC is a gas ionization detector made up of a mild steel tube of dimension 600$\times$10$\times$10\,cm$^3$. The tubes are sealed at both ends and filled with a P10 gas mixture (90\% argon and 10\% methane) at a pressure of approximately 25\% above the local atmospheric pressure. In the center, a 100\,$\mu$m thick tungsten wire is laid, serving as an anode, while the metal body is the cathode. The anode and cathode are electrically isolated using glass-to-metal hermetic seal. Each PRC is electrically biased with approximately, 3000\,V$_{\text{DC}}$. 58 PRCs are tightly placed in each layer. Alternate layers are placed orthogonally to each other and sandwiched with 15\,cm thick concrete slabs in between. Above the top-most layer, a 2\,m thick concrete absorber ($\sim$550\,g\,cm$^{-2}$) in an inverted pyramidal shape with an inclination of 45$^\circ$ ensures the filtering of soft components and provides an energy threshold of 1\,GeV for vertical muons, and sec($\theta$)\,GeV for muons incident at an angle $\theta$. 

\begin{figure}[t]
    \centering
    \includegraphics[width=0.85\textwidth]{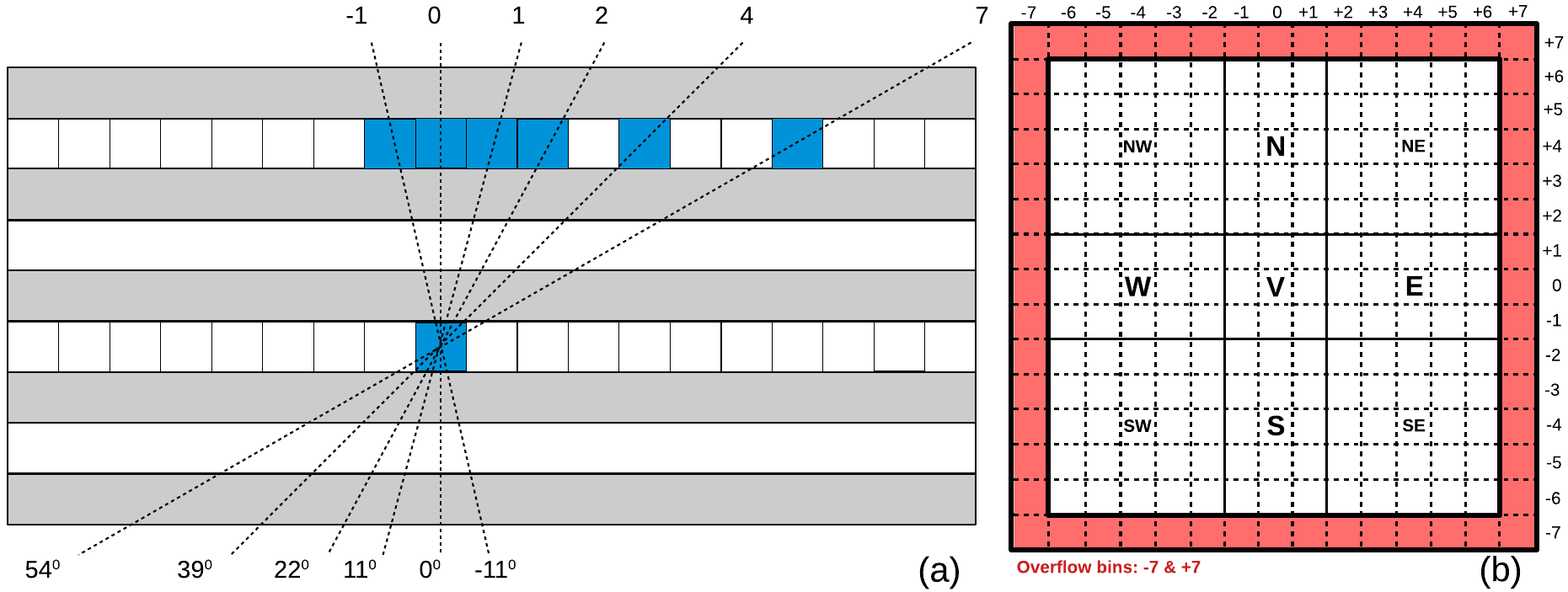}
    \caption{Representation of (a) muon direction reconstruction in a single projection using two layers, and (b) map of reconstructed 225 directions using four layers of PRCs. Solid lines represent the division into a coarser set of 9 directions after excluding the outermost overflow directions.}
    \label{Fig_01}
\end{figure}

Fig.\,\ref{Fig_01}a shows the reconstruction of the muon direction in a single projection using two layers of PRCs. Layers in the same projection (i.e. East-West or North-South) are separated by $\sim$52\,cm. For every muon that passes through all four layers of PRCs, the spatial separation of the hit PRC in the upper layer in a single projection with respect to the lower layer is calculated as shown in Fig.\,\ref{Fig_01}(a). The 45$^\circ$ inclined concrete absorber limits ($>$1\,GeV) the tracking of muons to $\pm$7 PRCs with respect to the impact of the bottom layer (15 directions in a single projection). Also, hits beyond $\pm$7 PRCs are binned into the $\pm$7th directions; thus, they are treated as overflow directions. When the directions derived from the two projections are combined, the direction of the muon can be reconstructed to 15$\times$15=225 directions covering 2.3\,sr sky (Fig.\,\ref{Fig_01}(b)). The DAQ software integrates the online reconstructed muons into these 225 directions as histograms for every 10 seconds of live time. It is also possible to combine the directions to make coarser directions as the analysis demands. Taking into account of the PRC width of 10\,cm and the 52\,cm layer separation in a single projection, this method of muon direction reconstruction provides a mean angular resolution of $\sim$4$^\circ$. The G3MT records about four billion muons every day. These are believed to be produced by CRs with energies more than 10\,GeV through EAS mechanishm.


\section{Causes of variations \label{causes}}

The G3MT consists of sixteen independent muon modules (M00--M15) each providing continuous measurements of the angular muon rates in 225 directions. It has been in continuous operation since 2000, making it an ideal choice for studying the long-term modulation of galactic cosmic rays (GCRs) and near-Earth phenomena. Since G3MT records about four billion muons daily, sudden changes as small as $\sim$0.002\% in the daily ($\sim$0.06\% per minute) muon flux can be detected significantly. However, the measured muon rate also shows spurious variations as a result of instrumental causes. 

\subsection{Systematic effects}

GCRs consist of high-energy particles, mainly protons and heavier nuclei, originating in outer space. When GCRs enter the heliosphere, they are modulated by the solar wind during their propagation in the interplanetary medium, causing them to be deflected. This leads to significant energy-dependent spatial and temporal variations in their intensity. Notably, the propagation of coronal mass ejections (CMEs) through the interplanetary medium, often accompanied by shocks, is a prime contributor to GCR variability. This variability encompasses phenomena such as precursor, Forbush decrease (FD), geomagnetic storm (GMS), and ground-level enhancement (GLE) \cite{1966P&SS...14..979Y, 2002ESASP.477..289K,Arunbabu_2015, 2015SoPh..290.1271B}. In addition, there are well-established periodic effects induced by the Sun at play. These include diurnal variation, 27-day solar rotation, 11-year solar cycle, and 22-year solar magnetic cycle \cite{1990SSRv...52..121V}. Each of these periodic phenomena induces analogous periodic responses in the observed intensity of GCRs. After being modulated by the solar magnetic field, GCRs reach the Earth's atmosphere, where they initiate EASs, producing a cascade of secondary particles. These secondaries travel several kilometers through the atmosphere during EAS development, making their flux highly sensitive to even small atmospheric changes. Consequently, variations in atmospheric parameters such as temperature and pressure directly influence the observed muon intensity \cite{Arunbabu_2017, Mohanty_2016_1}.

\subsection{Instrumental effects}

\begin{figure}
    \centering
    \includegraphics[width=0.85\textwidth]{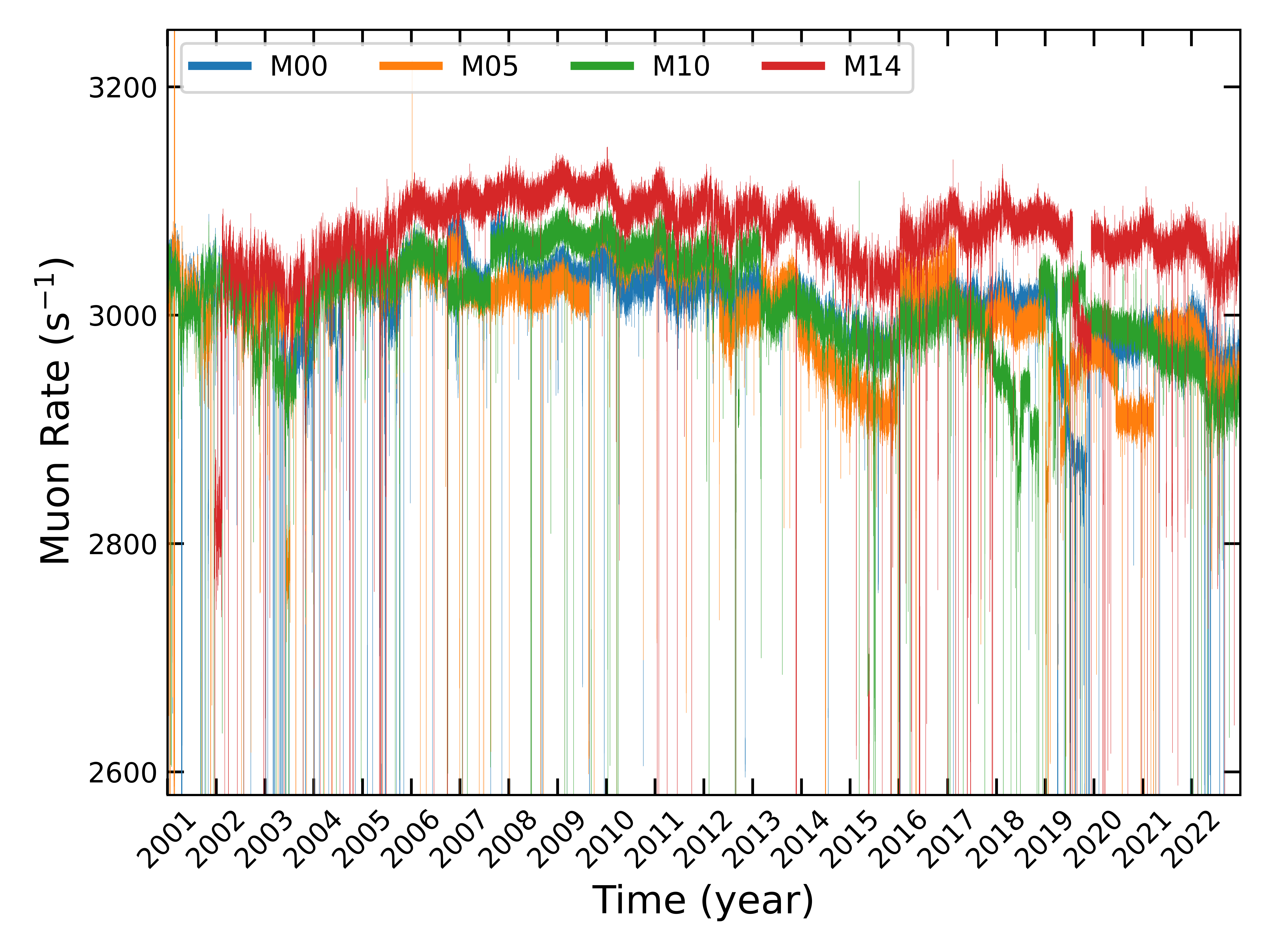}
    \caption{Sum of muon rates in all directions for four different modules during 2001-2022. Vertical drops are due to short breaks in data acquisition in individual modules. Short-term variations as well as a slow decline in the muon rate of certain individual modules can be observed, presumably due to short and long-term instrumental effects as well as variations in the muon flux.}
    \label{Fig_02}
\end{figure}

Despite diligent efforts to keep the detector in optimal working condition, sudden perturbations are observed in the muon rates. These perturbations are mostly short-term and are typically attributed to issues associated with the PRC, signal-processing electronics, and the DAQ \cite{Ramesh_2023}, such as sudden and momentary changes in the high voltage supplied to the PRC,  malfunction of the DC block capacitor, integrated amplifier-discriminator card, and various signal processing circuits. Although many of these issues last merely a fraction of a second, the measured muon rate is affected for at least 10 seconds, since the DAQ software integrates the angular muon rate in 10-second data packets. Minor leakages of P10 gas from the PRCs cause variations in detector efficiency over longer timescales and cause relatively small variations in muon rates over a long period of time. These variations are often not immediately noticed due to the presence of 12-hour pressure oscillations of much larger amplitude. Over several months, the pressure of the P10 gas within the PRC drops to that of the ambient atmosphere, enabling the air to penetrate into the PRCs. Once this happens, the data exhibit noticeable noise, prompting the manual disconnection of the PRCs for maintenance, leading to jumps in the recorded muon rate. Once we have a sufficient number of PRCs with low and/or zero rates in a given module, those are refilled during a planned shutdown. Subsequently, we observe a sudden increase in the muon rate of that module. Also, it should be noted that this effect is different among the modules. Fig.\,\ref{Fig_02} shows the uncorrected all-direction muon rates of four different modules collected during 2001--2022. The transient instrumental effects of data gaps and sudden variations, as well as a slow decline in the detector efficiency, can be observed. Although the G3MT can resolve the muon directions into 225 directions in each module, we present the all-direction muon rates for the subsequent discussions here. However, all the corrections can also be applied to individual directions.


\begin{figure}[t]
    \centering
    \includegraphics[width=0.99\textwidth]{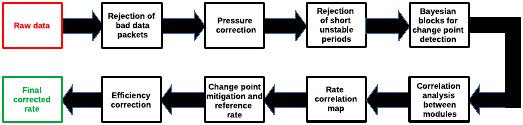}
    \caption{The flowchart represents various stages of identification and removal of instrumental effects.}
    \label{Fig_03}
\end{figure}

\section{Methodology \label{method}}

In this section, we discuss various instrumental effects and respective corrective measures with the final objective of efficiency correction. The identification and removal of instrumental effects are categorized into several steps consisting of, (i) rejection of bad data packets, (ii) pressure correction, (iii) rejection of short unstable periods, (iv) Bayesian blocks for change point detection, (v) correlation analysis between modules to identify reference modules, (vi) creation of a rate correlation map, (vii) change point mitigation and construction of the reference rate, and (viii) efficiency correction, shown  as flow chart in Fig.\ref{Fig_03} for the reader's convenience. Each of these aspects is discussed in the following subsections. 

\subsection{Rejection of bad data packets}

\begin{figure}[t]
    \centering
    \includegraphics[width=0.90\textwidth]{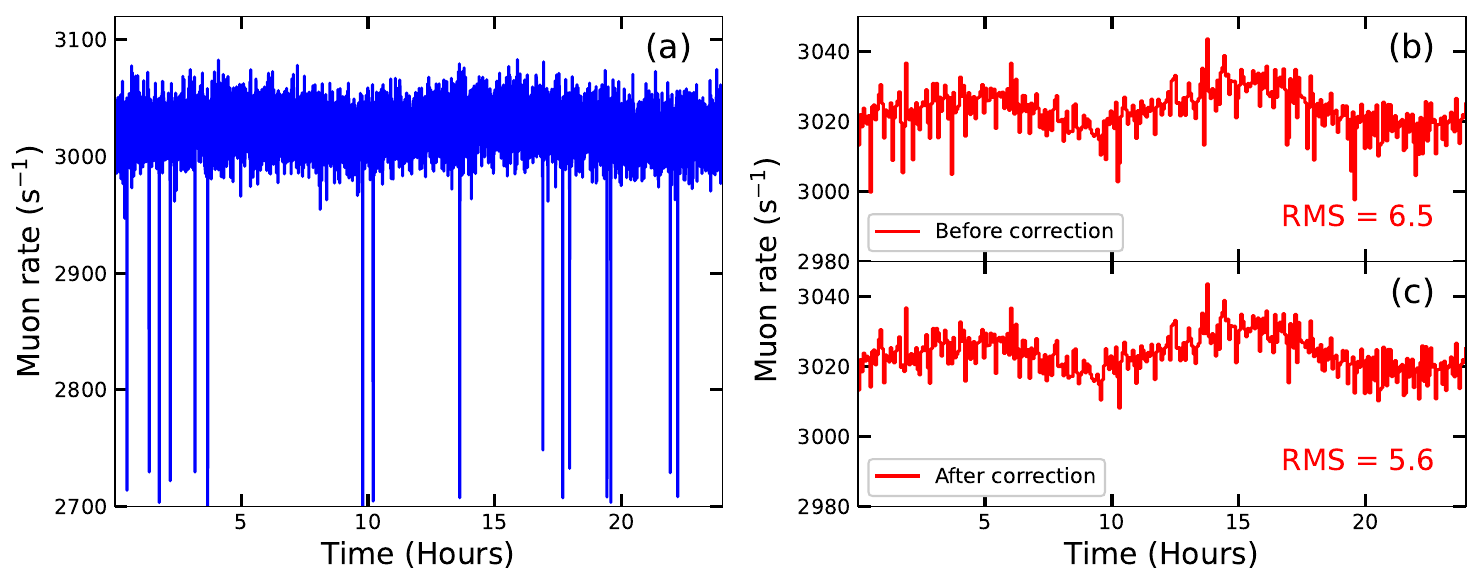}
    \caption{Variation of muon rate for M12 recorded on 3 January 2011 is shown in (a) for 10-second binning, that shows the sudden drops in the data packets. Figures (b) and (c) show the variation of muon rates for the same module in 4-minute binning. For each 4-minute bin, the data packets with the total muon rate beyond $\pm$3$\sigma$ are removed to exclude the bad data packets, and plotted in (c) that shows relatively lower fluctuation leads to reduced RMS, compared to (b).}
    \label{Fig_04}
\end{figure}

We devised and implemented a data rejection procedure in order to address a few instrumental effects mentioned in the previous section. As discussed in Sec.\,\ref{g3mt}, the G3MT records the muon rates in 10-second live data packets, which actually takes about 12 seconds to complete, due to the dead-time in the system. Whenever there is a disturbance in the high-voltage supplied to the PRC due to the trapping of moisture in the supply lines or the electrocution of insects, the response of the PRC changes, subsequently biasing the muon direction reconstruction. Even if this lasts for a fraction of a second, the entire data packet integrated over 10 seconds is affected  and tagged as a bad data packet, which is inevitable in the current data recording system. Fig.\,\ref{Fig_04}(a) shows muon rate variation of M12 recorded on 3 January 2011 for 10-second binning. The random occurrence of sudden drops in the rate is attributed to these issues. In such cases, even when the data is binned for a longer time interval (i.e., typically 4-minute binning in GRAPES-3, which corresponds to 1$^\circ$ of Earth's rotation), the bad data packet contaminates the true variation (Ref. Fig.\,\ref{Fig_04}(b)). We identified and removed such bad data packets by assuming that the muon rates follow Poissonian statistics. In each 4-minute aggregated time bin, ideally, one expects 24 data packets. However, due to the dead time of the system, only 20 data packets are produced. The bad data packets are identified with a $\pm$3$\sigma$ cut to the all-direction muon rate distribution of these data packets and rejected from further analysis. This exercise efficiently rejects such bad data packets and the resultant muon rate variation is shown in Fig.\,\ref{Fig_04}(c), with the reduced RMS of 5.6, compared to the RMS of 6.5 obtained before removing the bad data packets. 

\subsection{Pressure correction}

\begin{figure}[t]
    \centering
    \includegraphics[width=0.70\textwidth]{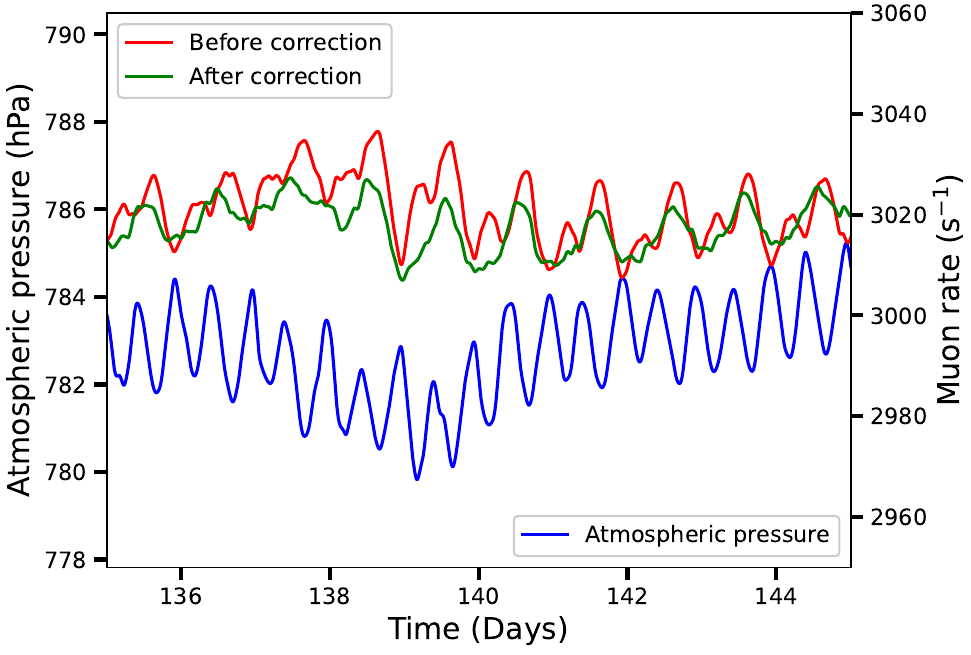}
    \caption{The blue line represents atmospheric pressure, while the red and green lines illustrate the muon rates of M00 before and after undergoing pressure correction for the period from 15--25 May 2010. A one-dimensional Gaussian filter with $\sigma$=10 was applied for better visualization, but the original data was used in the actual analysis.}
    \label{Fig_05}
\end{figure}

Atmospheric pressure plays a crucial role in modulating the muon flux. It is influenced by the heating of the Earth's atmosphere by the Sun, which leads to diurnal and semi-diurnal variations in the observed pressure. The higher the atmospheric pressure, the higher the probability of muon interactions and decay in the atmosphere, and vice versa. As a result, the atmospheric pressure is expected to be anti-correlated to the observed muon rate. A previous study by GRAPES-3 showed a pressure coefficient ($\beta$) of (--0.128$\pm$0.005)\% hPa$^{-1}$ \cite{Mohanty_2016_1}. In this study, removing these short-term systematic effects was necessary before addressing instrumental effects, as the latter are independent and can be long-term. This systematic variation in the measured muon rate can be corrected using the method discussed in~\cite{Mohanty_2016_1}. The correction for atmospheric pressure effects can be applied at any stage of data processing. However, in some studies, analyzing pressure-uncorrected muon data is essential, such as for detecting atmospheric waves generated by the Hunga Tonga-Hunga Ha’apai volcanic eruption \cite{Hari_2023}. Fig.\,\ref{Fig_05} illustrates variations in muon flux corresponding to changes in atmospheric pressure, with a 12-hour periodicity. For better visualization, we applied a one-dimensional Gaussian filter with a standard deviation ($\sigma$) of 10 to both the atmospheric pressure and muon rate data. However, in the actual analysis, we used the original data without smoothing. The blue line represents atmospheric pressure, the red line depicts the muon rate recorded by M00 from 15--25 May 2010, and the green dotted line showcases the muon rate after applying pressure correction. To achieve this correction, we employed Eq.\,\ref{eq:1}:

\begin{equation}\label{eq:1}
    \text{R}_{\text{corr}}(\text{t}) = \frac{\text{R}_{\text{obs}}(\text{t})}{(1+\beta \Delta \text{P})}
\end{equation}

wherein, $\beta$ was set at --0.128\% hPa$^{-1}$. This pressure-corrected data, displayed by the green dotted line, provides a clear depiction of the solar diurnal variations resulting from the Earth’s rotation.

\subsection{Rejection of short unstable periods}

\begin{figure}[t]
    \centering
    \includegraphics[width=0.90\textwidth]{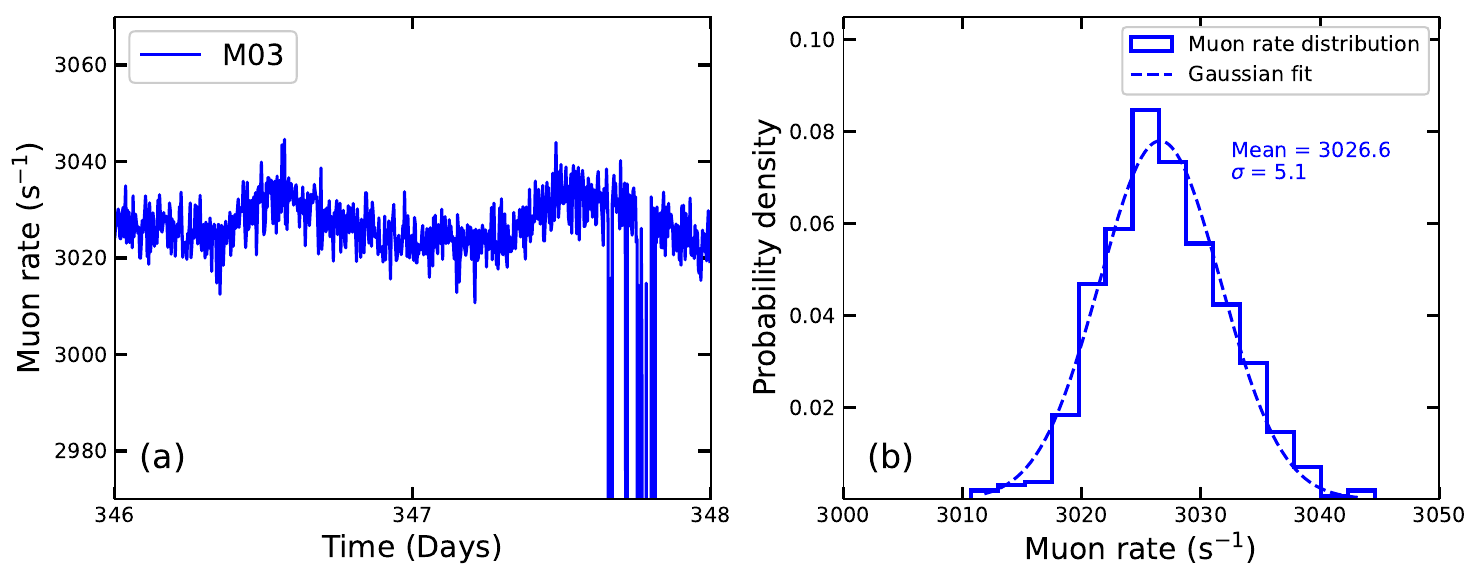}
    \caption{(a) Muon rate variation of M03 during 13--14 December 2005, and (b) a Gaussian fit to the muon rate distribution for the same period.}
    \label{Fig_06}
\end{figure}

\begin{figure}[t]
    \centering
    \includegraphics[width=0.95\textwidth]{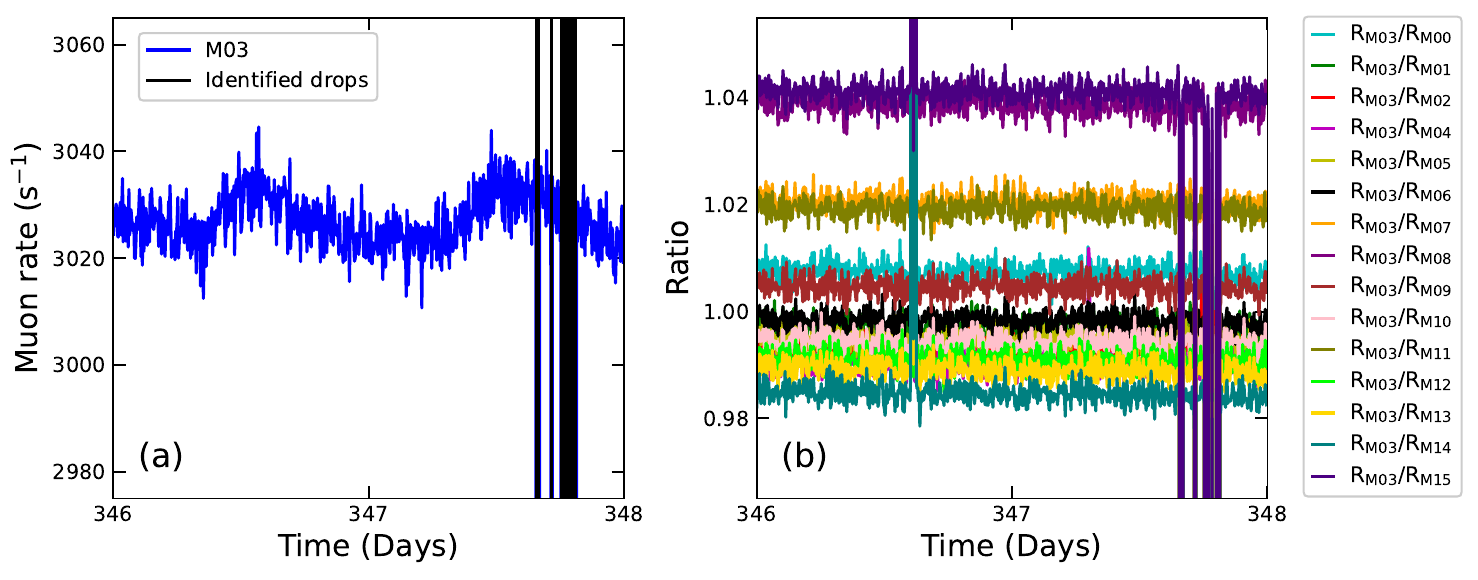}
    \caption{(a) Periods with sudden drops in the muon rate are indicated by the black shaded bands after applying the cut $\pm$4$\sigma$ between 13--14 December 2005 for M03, and (b) ratio of muon rate of M03 with the other 15 modules during the same period.}
    \label{Fig_07}
\end{figure}

Changes in muon rate that last less than a day were identified by fitting a Gaussian function to the daily muon rate. Fig.\,\ref{Fig_06}(a) illustrates the variation of muon rate in M03 from 13--14 December 2005, during which we observed drops in muon rate on 14 December 2005. The rate distribution is fitted with a Gaussian function (Fig.\,\ref{Fig_06}(b)), any significant deviations from the fit are rejected to exclude bad data within each day. The fit yielded a mean of 3026.6 counts and a standard deviation of 5.1 counts. Muon rates that fall beyond $\pm$4$\sigma$ on the mean rate are treated as sudden fluctuation. We chose to fit the data over a two-day period, since the daily fluctuations are higher during high solar activity. 

\begin{figure}[t]
    \centering
    \includegraphics[width=0.90\textwidth]{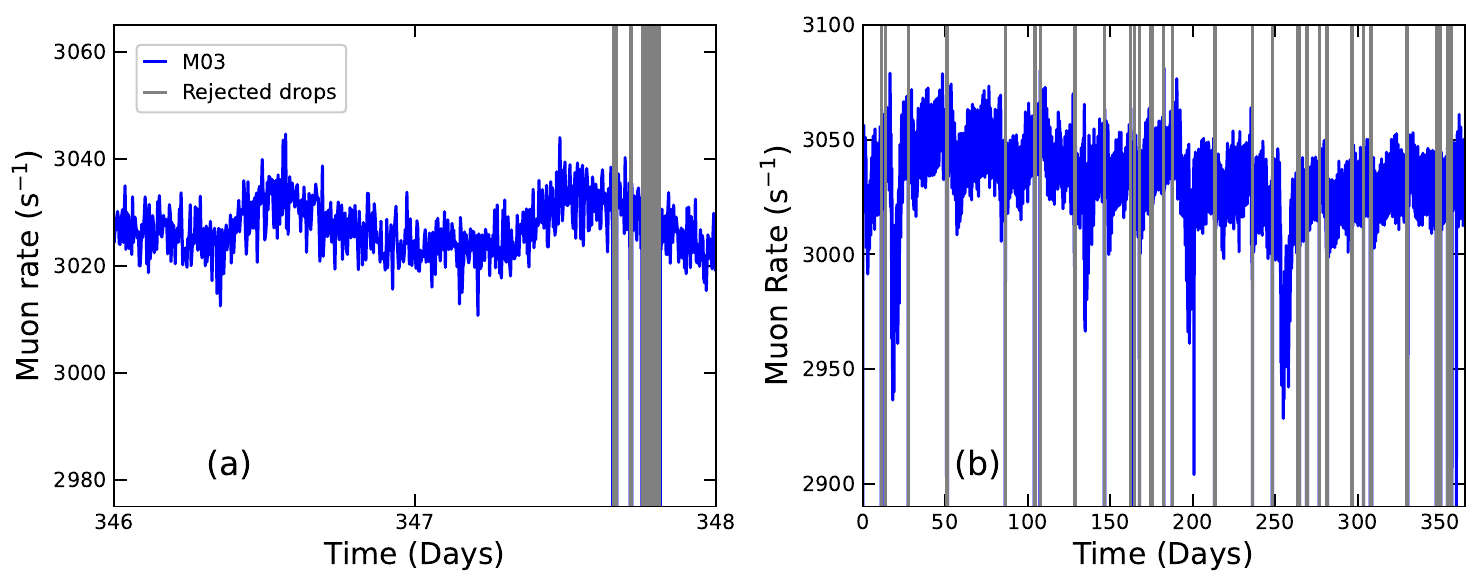}
    \caption{Rejected data periods of M03 indicated by gray lines during (a) 13--14 December 2005, and (b) the year 2005.}
    \label{Fig_08}
\end{figure}

Fig.\,\ref{Fig_07}(a) displays sudden drops in the muon rate of M03 after applying the $\pm$4$\sigma$ cut, indicated by the black line. However, it is unclear whether these drops were caused by a detector problem or some physical phenomenon. As the variation in muon rate due to physical phenomena is common to all 16 modules, to eliminate these effects, we can calculate the ratio with the rate from the other modules, leaving us with only variations caused by detector problems. Fig.\,\ref{Fig_07}(b) displays the ratio of muon rate of M03 with the remaining 15 modules during the period 13--14 December 2005. A flat trend among the ratios is indicative of the removal of physical phenomena that are common to all the muon modules. However, drops corresponding to a few muon modules or super-modules due to instrumental effects can be identified and added to the rejection list based on this quantitative selection. To account for the fact that drops can occur in multiple modules during the same period, we apply a 70\% cut. Fig\,\ref{Fig_08}(a) and (b) show the rejection list of M03 during the period 13--14 December 2005 and for the whole year 2005, respectively. Only $\sim$1.47\% of the total duration is identified as bad periods and added to the rejection list, as shown in Fig.\,\ref{Fig_08}(b).

\begin{figure}[t]
    \centering
    \includegraphics[width=0.98\textwidth]{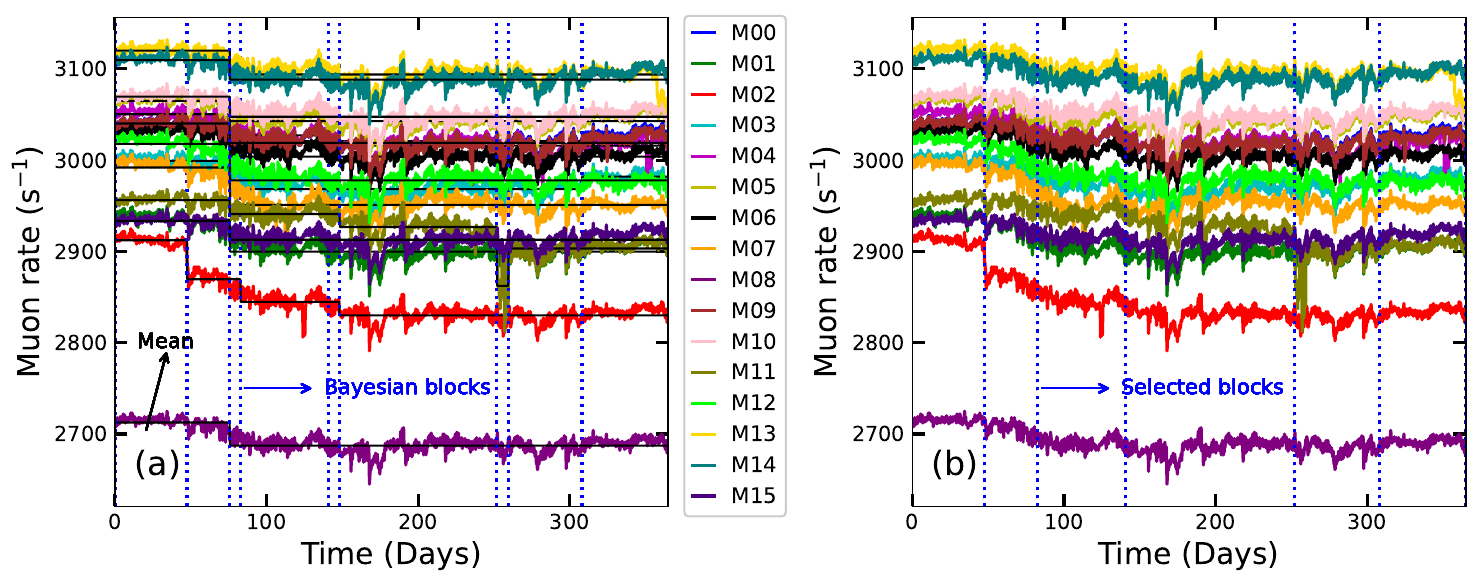}
    \caption{Muon rate profiles of all 16 modules for the year 2011. The blue dotted lines separate the identified Bayesian blocks (a) before and (b) after applying two cuts: (1) removal of simultaneous blocks appearing in all 16 modules, and (2) merging of change points separated by less than 30 days. The total number of blocks is reduced from 71 in (a) to 6 in (b). The black solid lines in (a) are the mean rates of the muon modules within each Bayesian block.}
    \label{Fig_09}
\end{figure}

\begin{figure}[t]
    \centering
    \includegraphics[width=0.40\textwidth]{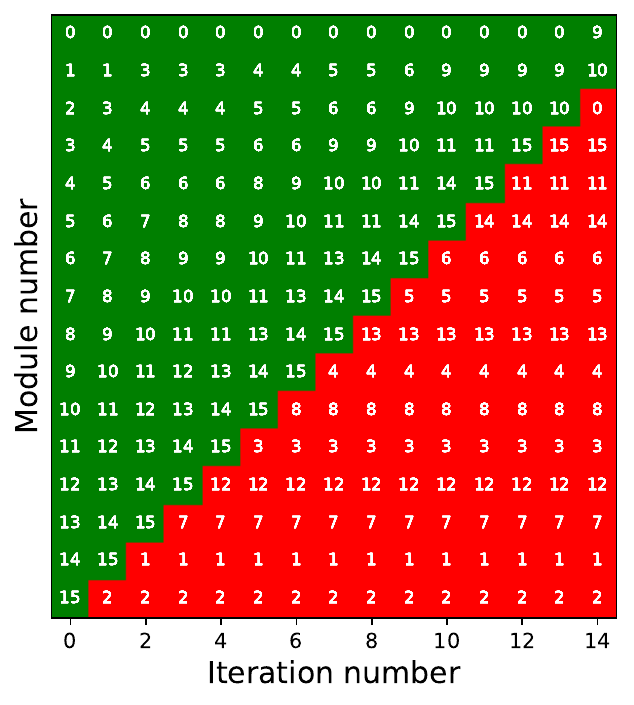}
    \caption{Matrix of module and iteration numbers illustrating the iterative determination of a stable pair of modules. The module with the lowest average correlation coefficent with the remaining modules is eliminated at each iteration. M09 and M10 were found to be the stable modules for the first Bayesian block in 2011.}
    \label{Fig_10}
\end{figure}

\subsection{Bayesian blocks for change point detection \label{longterm}}

The variation in G3MT's efficiency due to gas leakages of PRCs can persist for extended periods, depending on the rate of gas leakage and the duration of preventive maintenance. Since each muon module comprises 232 PRCs, it is impractical to designate a single module as consistently stable over an extended period due to the inherent susceptibility of PRCs to gas leakages. We use the Bayesian blocks algorithm to discretize the time series of rates of all 16 modules into periods. Bayesian block is a valuable tool for time series data analysis, offering an adaptive and data-driven approach for temporal binning, ensuring that the structure and significance of the data are properly preserved. The algorithm determines if periods of data are compatible with each other using a threshold called p-value, and breaks the period into blocks. This makes Bayesian blocks particularly well suited for uncovering hidden patterns, detecting change points, and extracting meaningful information from time series datasets. Fig.\,\ref{Fig_09}(a) shows the muon rates of all 16 modules during the year 2011. The change points identified in each module using Bayesian blocks with a p-value of 0.01 are indicated using blue dotted lines. The black solid lines indicate the mean rates of the muon modules within each block. In total, we found 71 blocks from all 16 modules ($\sim$5 blocks per module). Fig.\,\ref{Fig_09}(b) shows the final set of Bayesian blocks, selected after merging change points separated by less than 30 days, leaving behind six blocks, spanning 48, 35, 58, 111, 56 and 57 days, respectively. 

\subsection{Correlation analysis between modules}

\begin{figure}[t]
    \centering
    \includegraphics[width=0.90\textwidth]{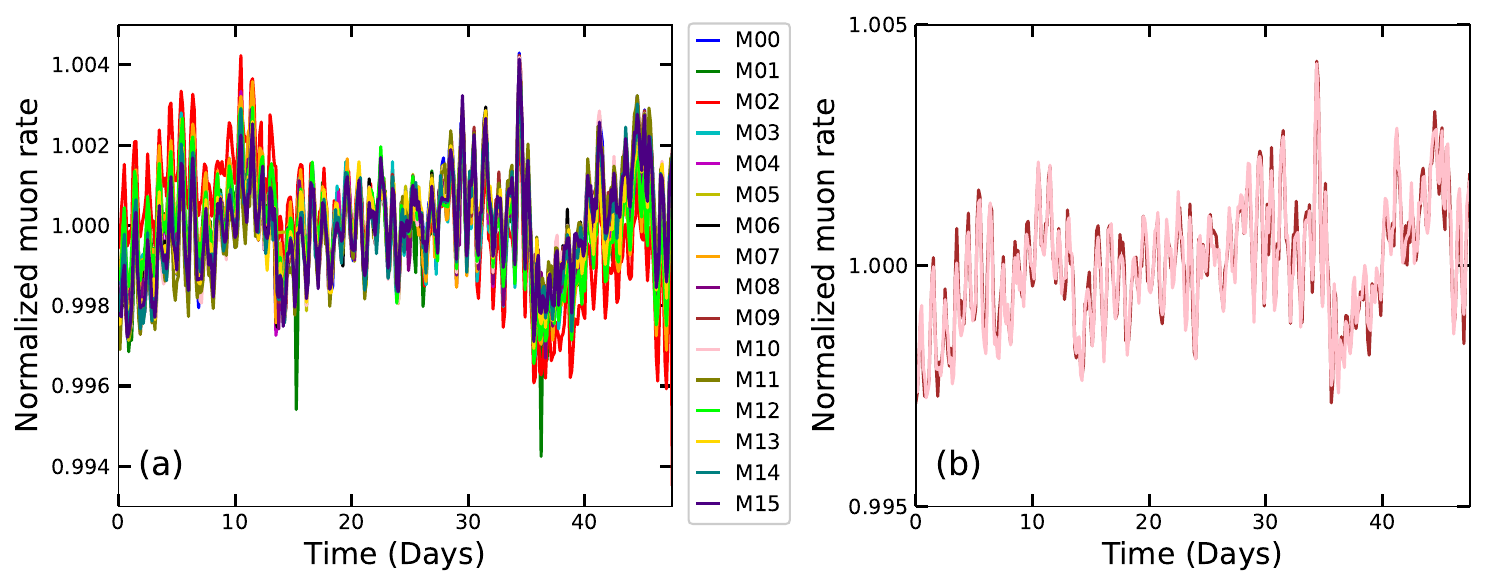}
    \caption{Normalized muon rates of (a) all 16 modules, and (b) modules 9 and 10 for the first Bayesian block of the year 2011, obtained from Fig.\,\ref{Fig_10}(b).}
    \label{Fig_11}
\end{figure}

We calculate the correlation coefficient (CC) of each module with respect to the remaining 15 modules, and the average of these 15 CCs is computed. The module with the lowest average CC among the 16 is considered highly affected by efficiency variations. Consequently, we remove that module and repeat this exercise for the remaining 15 modules and the corresponding 14 possible pairs. This process is repeated until the two most correlated modules are remaining. These are treated as the best reference modules for the specific Bayesian block in consideration. This exercise resulted in M09 and M10 being identified as the most stable, during the first Bayesian block in the year 2011. This is illustrated in Fig.\,\ref{Fig_10}, which presents a matrix of module and iteration numbers that illustrate the convergence to the stable modules at the last iteration. The muon rates in the first Bayesian block for the year 2011 are shown in Fig.\,\ref{Fig_11}(a) and (b) for all modules and M09 and M10, respectively. We can clearly see that M02 displays the highest level of efficiency variation, whereas M09 and M10 are the lowest, which makes them the most stable modules. 

\begin{figure}[t]
    \centering
    \includegraphics[width=0.50\textwidth]{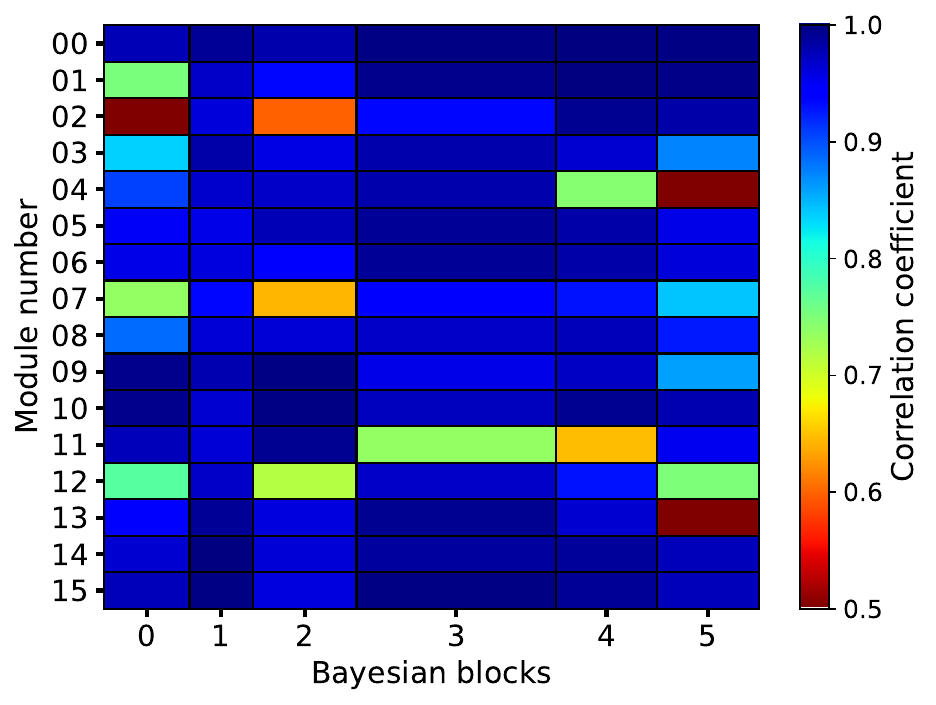}
    \caption{Rate stability map of all muon modules in the different identified Bayesian blocks for the year 2011. The colours correspond to the average of the correlation coefficients of that module to the other 15 modules.}
    \label{Fig_12}
\end{figure}

\subsection{Rate correlation map}

\begin{figure}[t]
    \centering
    \includegraphics[width=0.90\textwidth]{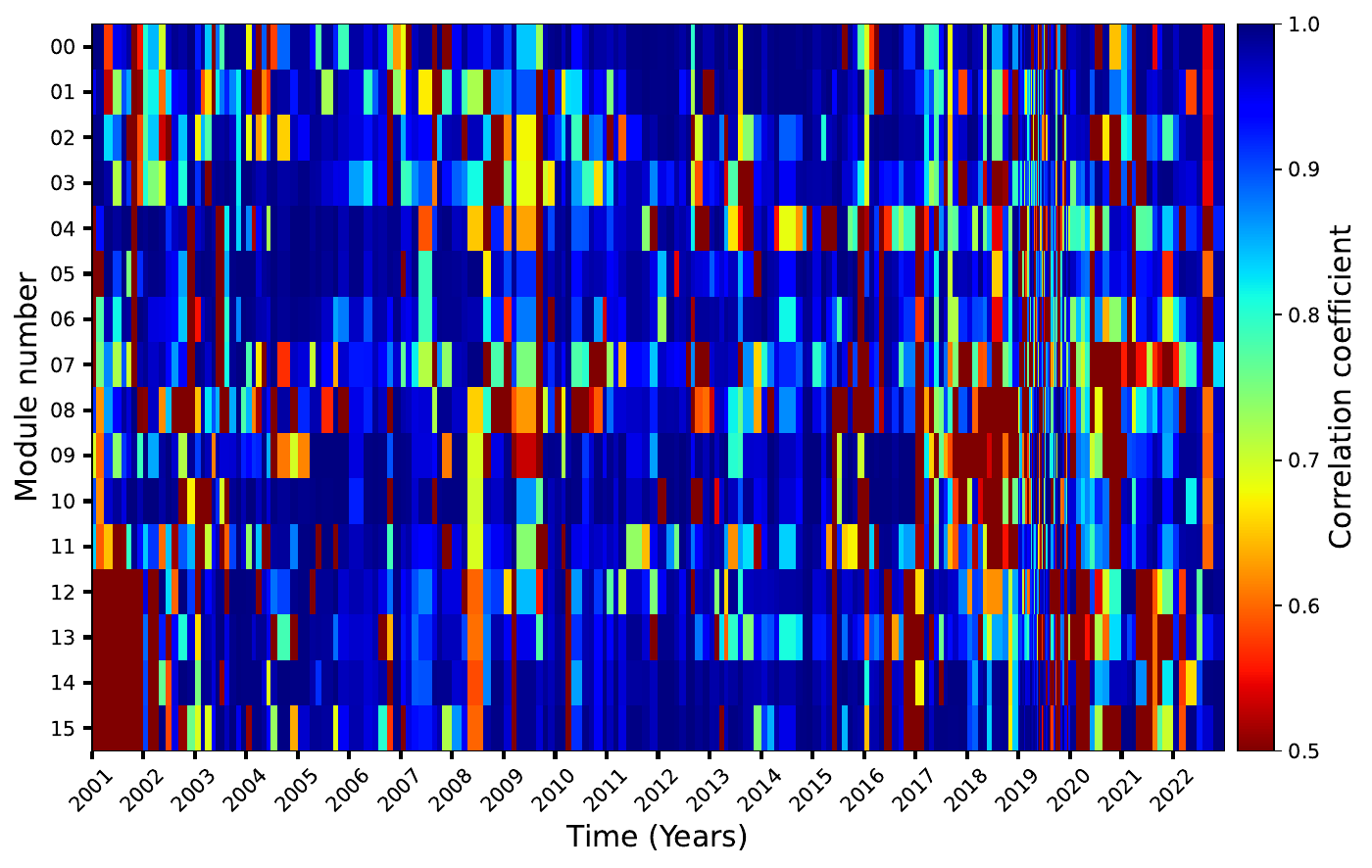}
    \caption{Rate stability map of all muon modules in the different identified Bayesian blocks for the year 2001-2022. The colours correspond to the average of the correlation coefficients of that module to the other 15 modules.}
    \label{Fig_13}
\end{figure}

Once the two most stable modules are identified, the average rate of those two modules is used as a reference for further analysis. This is in contrast with the previous method which relied on a reference module identified by the experiment operator \cite{Mohanty:2017umr}. The CC of all 16 module rates with reference to the average rate of stable modules is calculated for each Bayesian block to derive the rate stability map for the entire year. Fig.\,\ref{Fig_12} shows the rate stability map of the year 2011 obtained from this exercise. We extend this exercise for the entire duration of 22 years (2001--2022), shown in Fig.\,\ref{Fig_13}. These two figures show the muon rate stability map in terms of color contours. Shades of blue along the horizontal axis indicate the high stability of its efficiency over an extended period of time. Overall, the muon modules have been found to be stable over the past 22 years, except for the latter years during 2018--2020. During the years 2018 and 2019, the G3MT modules underwent an upgrade of their front-end electronics \cite{Ramesh_2023}, and further, Covid-19 restricted the preventive maintenance of G3MT modules in 2020. 

\subsection{Change point mitigation and reference rate}

\begin{figure}[t]
    \centering
    \includegraphics[width=0.90\textwidth]{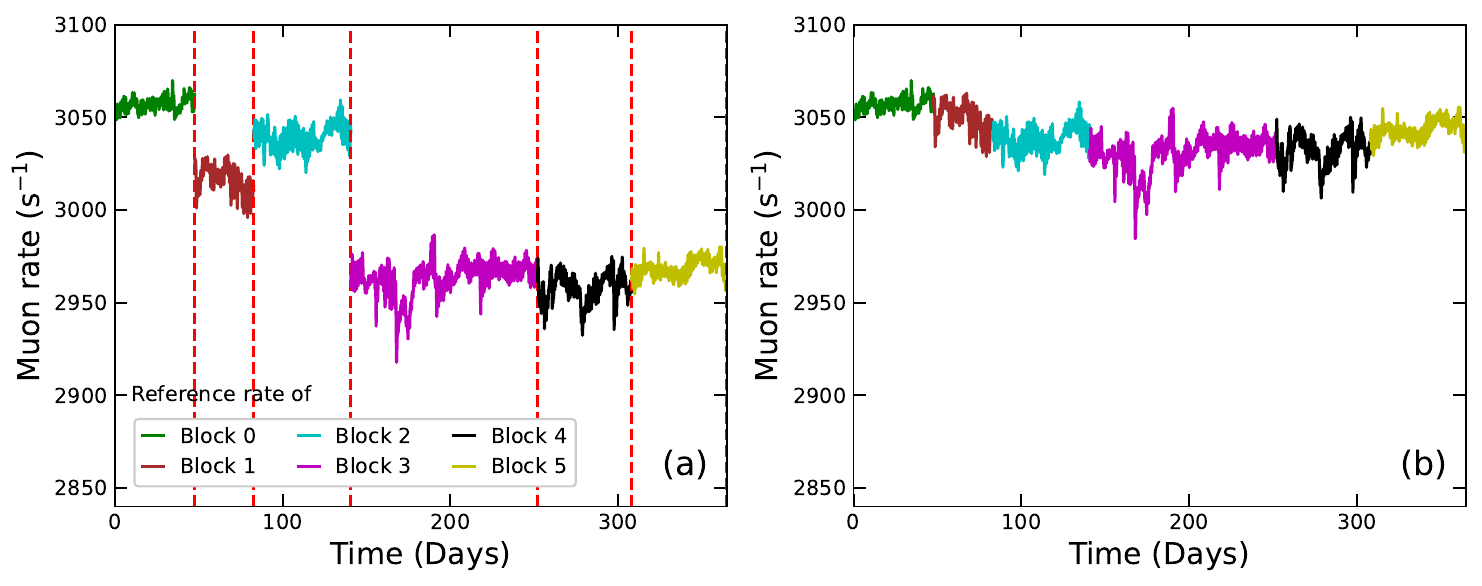}
    \caption{Variation of (a) reference muon rates based on the average of stable module rates in each Bayesian block before data stitching, and (b) muon rate after data stitching for the year 2011.}
    \label{Fig_14}
\end{figure}

\begin{figure}[t]
   \centering
    \includegraphics[width=0.90\textwidth]{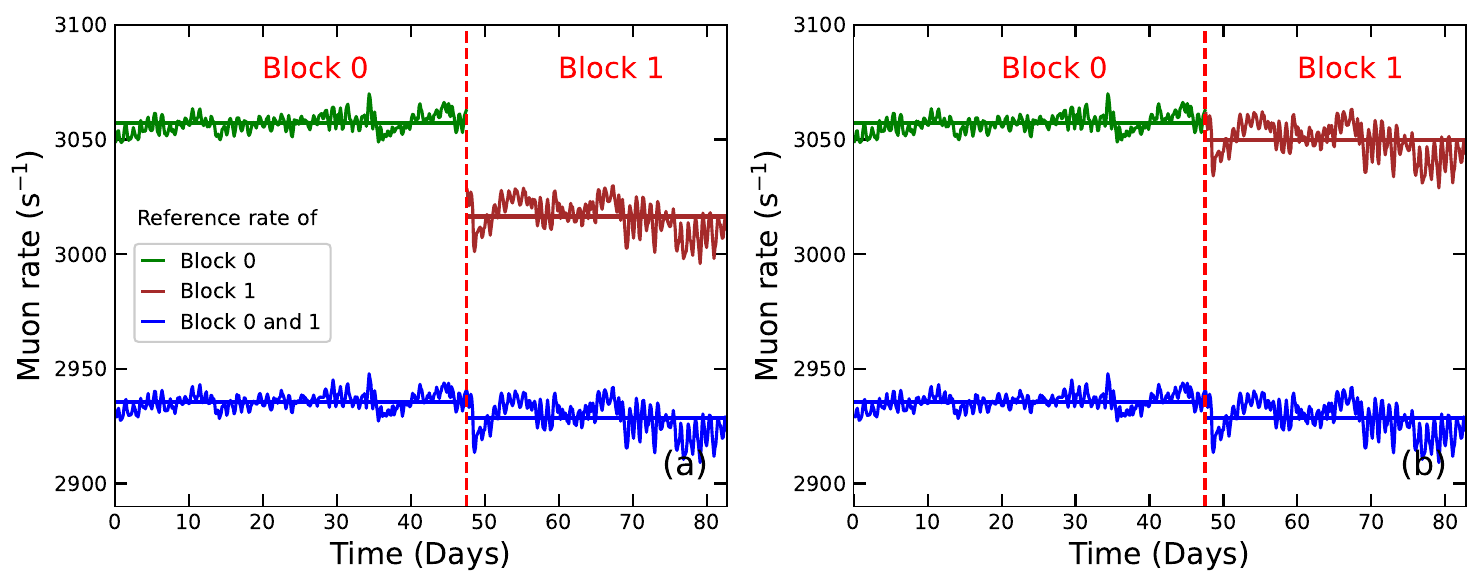}
    \caption{Representation of data stitching process of two Bayesian blocks in which M00 (shown in blue) is the common reference. The average rates of Bayesian blocks 0 and 1 (a) before, and (b) after data stitching for the year 2011.}
    \label{Fig_15}
\end{figure}

The previous steps provide the most stable modules in each Bayesian block. However, these are not necessarily the same throughout the whole duration. Fig.\,\ref{Fig_14}(a) shows the Bayesian blocks of the year 2011 with stable module rates. As can be seen, the baselines of these six blocks are different, which implies the identification of different stable modules in the process. In order to further correct the modules' efficiency variation for a period, a common reference module needs to be identified, which in turn necessitates a common baseline. Thus, the next step of data stitching is used to adjust the baselines of all Bayesian blocks to a common baseline. This is achieved using the rate stability map. We select the module with the highest average CC in two consecutive Bayesian blocks and use it for the data stitching of those two blocks, mitigating the change point between the two blocks. Fig.\,\ref{Fig_15}(a) depicts one such scenario in which M00, identified based on the rate stability map in Fig.\,\ref{Fig_12} is used as a baseline reference for data stitching of the first two Bayesian blocks 0 and 1 for the year 2011. Subsequently, the baseline corrected rates of these two Bayesian blocks are shown in Fig.\,\ref{Fig_15}(b) after data stitching. Similarly, Fig.\,\ref{Fig_14}(b) shows the outcome of data stitching for the year 2011. 

\subsection{Efficiency correction}

\begin{figure}[t]
    \centering
    \includegraphics[width=0.7\textwidth]{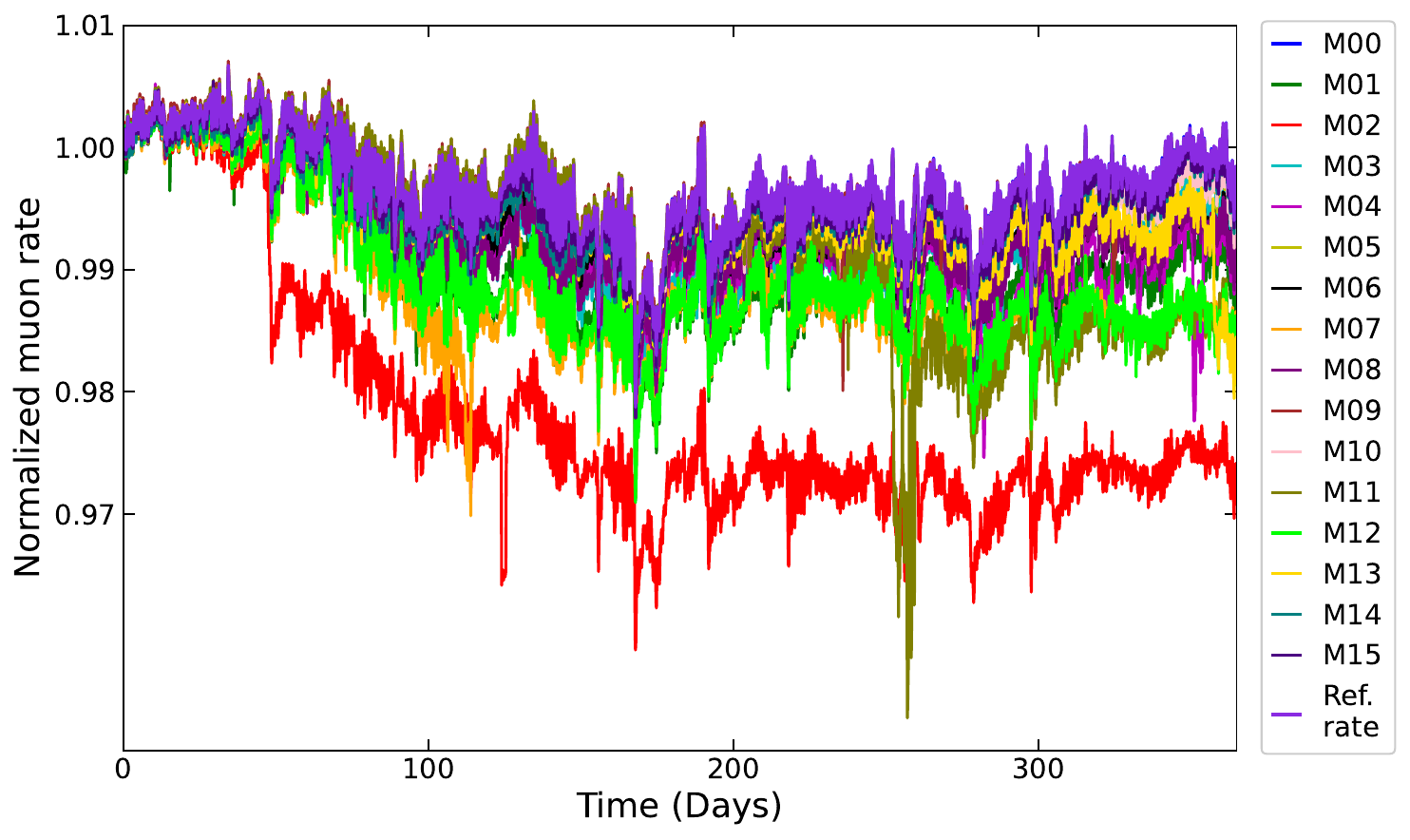}
    \caption{Muon rate ratio of 16 modules with respect to the reference rate for the year 2011.}
    \label{Fig_16}
\end{figure}

\begin{figure}[t]
   \centering
    \includegraphics[width=0.90\textwidth]{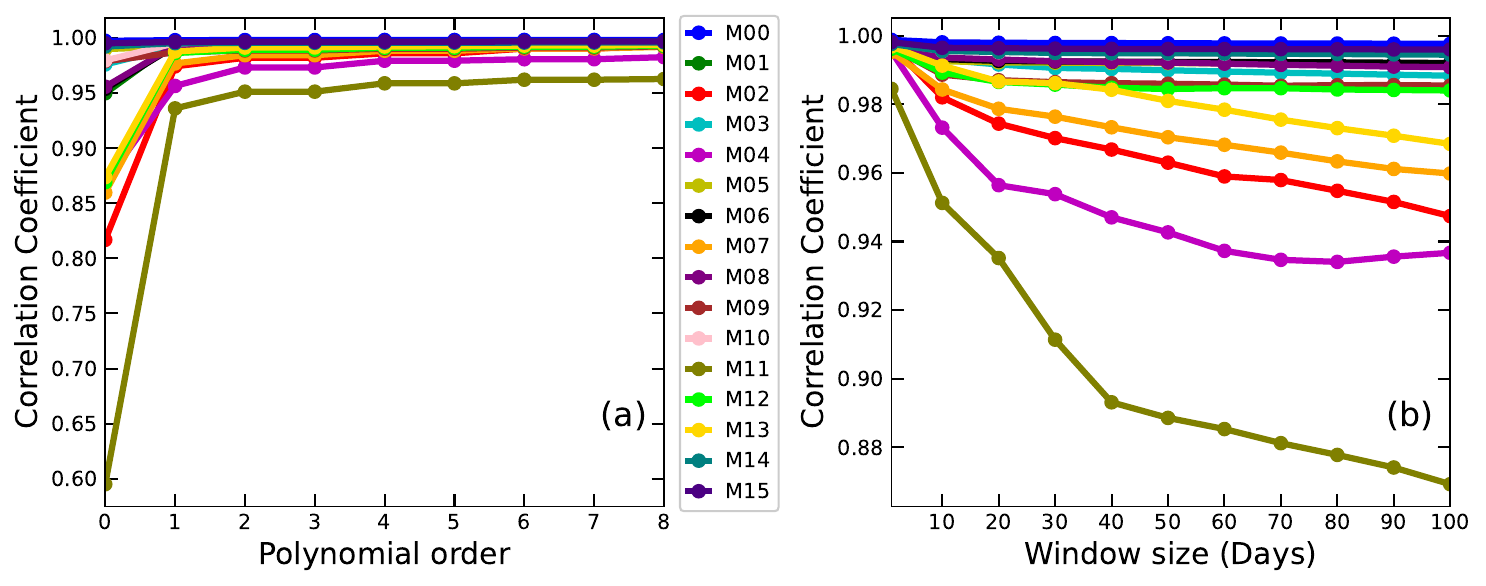}
    \caption{(a) Correlation coefficient of all 16 modules after efficiency correction with the reference module, calculated with a fixed window size of 10 days, as a function of polynomial order for the year 2011. The correlation coefficient converges at a polynomial order of 2. (b) Correlation coefficient as a function of window size, with a fixed polynomial order of 2. The correlation coefficient decreases as the window size increases, leading to the selection of a 10-day window for efficiency variation correction.}
    \label{Fig_17}
\end{figure}

\begin{figure}[t]
   \centering
    \includegraphics[width=0.90\textwidth]{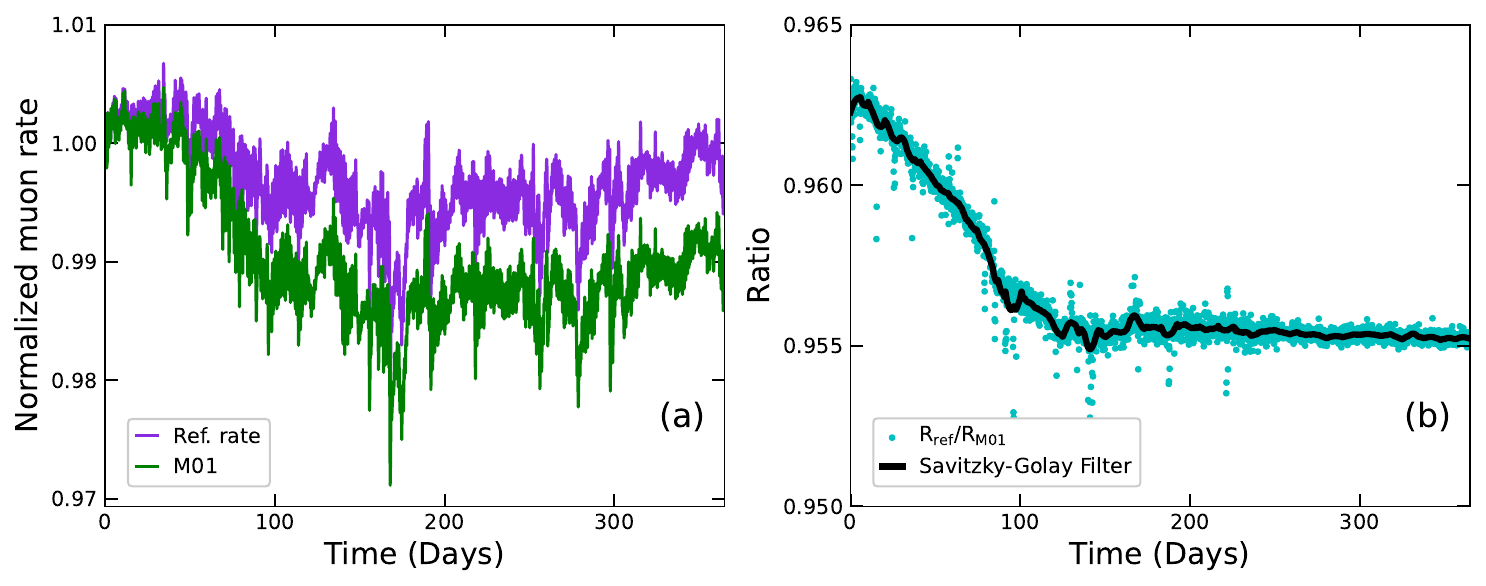}
    \caption{(a) Shows the variation of normalized muon rate of M01 in comparison with the reference rate for the year 2011. The declining rate of M01 due to efficiency variation can be seen clearly. (b) Modeling of the ratio of reference rate to M01 with SG filter.}
    \label{Fig_18}
\end{figure}

\begin{figure}[t]
   \centering
    \includegraphics[width=0.90\textwidth]{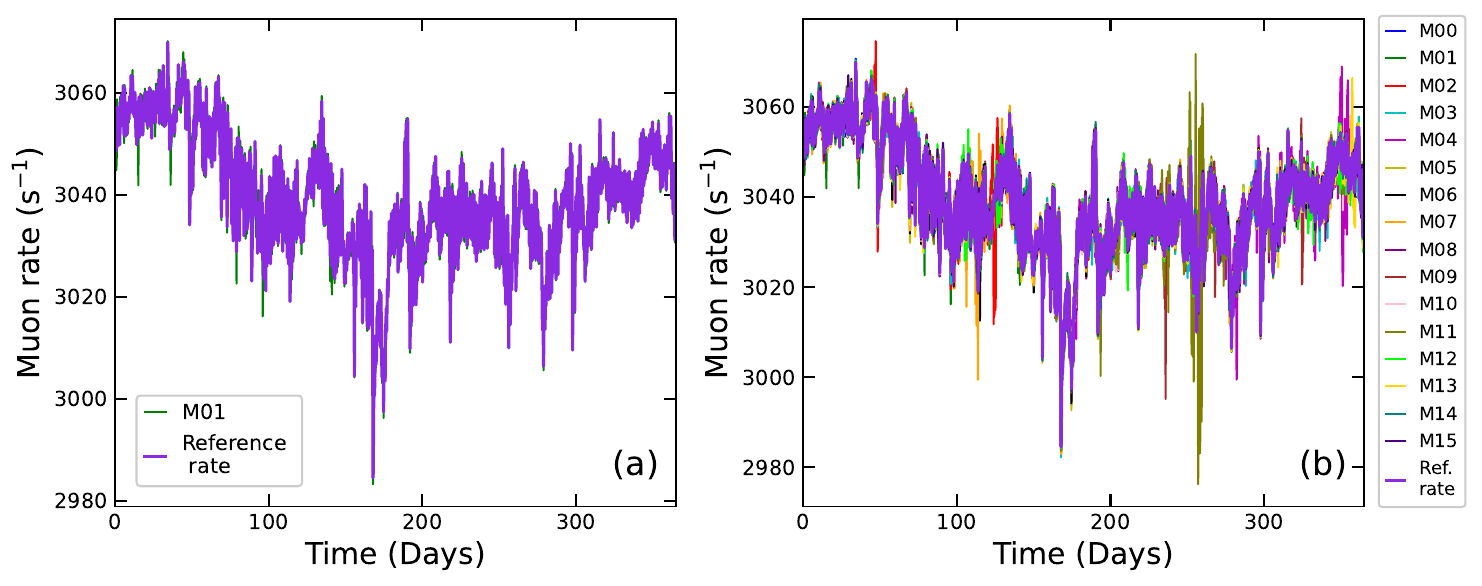}
    \caption{(a) the efficiency corrected muon rate of M01 using the Savitzky-Golay filter for the year 2011, and (b) for all 16 modules.}
    \label{Fig_19}
\end{figure}

The final step is to correct for the efficiency variation of muon modules, which is a long-term instrumental effect. Since each muon module is entirely independent, we can observe its efficiency variation by calculating the ratio with a reference module. Fig.\,\ref{Fig_16} shows the muon rate ratios of all 16 modules with respect to the reference rate obtained from the steps so far, for the year 2011. The module ratios are used for the characterization of efficiency variation. We employed a Python-based implementation of the Savitzky-Golay (SG) filter \cite{Savitzky-Golay}, which is based on a combination of running averages and higher order polynomial correction, to track each module's efficiency variation as a function of time. However, the SG filter needs a choice of two parameters to be specified, namely the window size and order of the polynomial, for fine-tuning the filter to model either slow or fast variations. To choose these parameters, we tested the variability of these two parameters on a wide range of values based on a convergence test. From these tests, we observe an asymptotic behavior on these two parameters at a window size of 10 days and polynomial order of 2 (Fig.\,\ref{Fig_17}). We adopt these values as optimum parameters of the SG filter in our method. This particular choice of parameters ensures that the slow variations are tracked by excluding local and transient fluctuations of individual modules, which makes the SG filter a valuable tool for understanding the time variation of the signal. Fig.\,\ref{Fig_18}(a) shows the variation of normalized muon rate of M01 in comparison with the reference rate, for the year 2011. The declining rate of M01 due to efficiency variation can be seen clearly. Subsequently, we model the ratio of the reference rate to the muon rate of M01 using SG filter with a window size of 10 days and polynomial order of 2 (Ref. Fig.\,\ref{Fig_18}(b)), The SG modeled parameters are further used to correct the efficiency variation of M01, as shown in Fig.\,\ref{Fig_19}(a). Subsequently, Fig.\,\ref{Fig_19}(b) shows the efficiency corrected muon rates of all 16 modules for the year 2011.

\begin{figure}[t]
    \centering
    \includegraphics[width=0.90\textwidth]{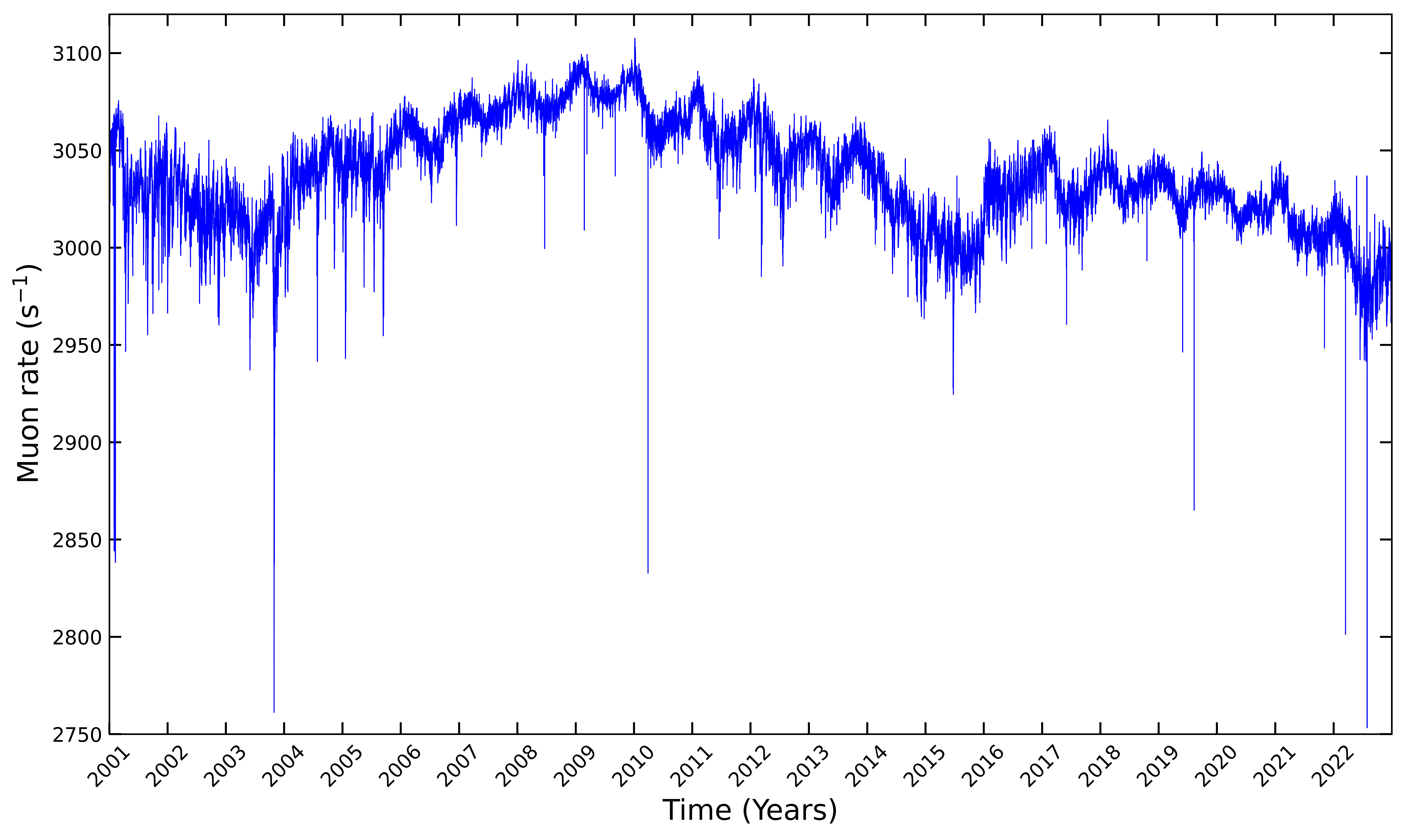}
    \caption{Variation of final efficiency corrected combined muon rate of all 16 modules and for all directions during the period 2001--2022. This rate is corrected for all the known systematic variations observed in the muon modules and is ideally suited for immediate physics analysis. The sudden drops seen in the data are due to the FD events. The largest drop seen in 2003 is associated with the biggest FD event recorded by the G3MT on 29 October 2003 with a recorded peak muon deficit of about 8\% \cite{Nonaka_2006}.}
    \label{Fig_20}
\end{figure}

\subsection{Final corrected muon rate}

As seen in Fig.\,\ref{Fig_19}(b), the muon rates are corrected for the slow variation borne out by the declining muon modules' efficiency. At this point, the module rates can be further combined into a single muon rate profile for all directions and/or individual directions. We present the combined all-direction muon rate variation of all the modules for the period 2001--2022 in Fig.\,\ref{Fig_20} after all the known instrumental effects are removed. The muon rate variation, which spans almost two solar cycles and a solar magnetic cycle, carries a lot of valuable information. As can be seen, there are numerous transient and slow variations. The sudden drops seen in the data are due to the FD events, triggered by the flaring activity in the Sun. One notable example is the largest drop seen during the year 2003, which is associated with one of the biggest FD event recorded by the G3MT on 29 October 2003, triggered by an X17 class solar flare that occurred on the previous day \cite{Nonaka_2006}. The G3MT recorded a deficit of about 8\% in the all-direction muon rate. The slow variations could be due to the other physical mechanisms of CRs during their propagation in the interplanetary medium and Earth's atmosphere. We reported one such effect of change in muon rate due to upper atmospheric temperature variation using six years of muon rate (2005--2010) \cite{Arunbabu_2017}. This study can be further extended to see the dependence during the different solar cycles using the data processed in this work. Also, it is possible to explore the dependence of muon rate variation as a function of interplanetary magnetic field.

\section{Validating the method}

\begin{figure}
    \centering
    \includegraphics[width=0.6\textwidth]{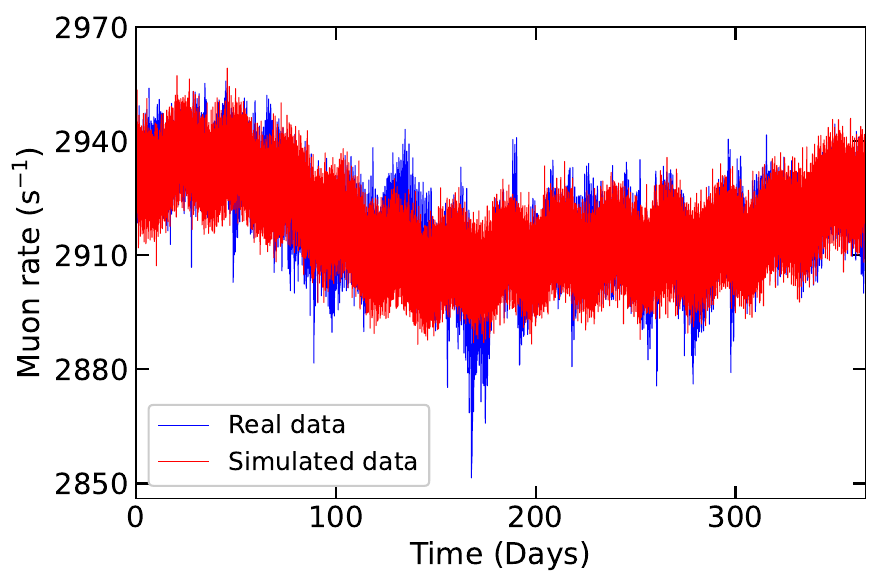}
    \caption{The blue line represents the variation of muon rates for M15 during 2011, and the red line shows the simulated data.}
    \label{Fig_21}
\end{figure}

We evaluate the performance of the method described hitherto, by comparing with the legacy method. To do so we simulate a mock muon rate modelled along real data. This is done both with a simulated mock muon rate, modelled on real data, and by using data from a neutron monitor as a proxy for the primary CR flux.

\subsection{Simulation of muon data}

For the simulation, we focus on the year 2011, during which M15 was identified as a relatively stable module based on logbook records. Thus, M15 was chosen as the reference for simulating muon rates. The simulation used mock data to replicate the periodic variations observed in real data, including: 1-day cycle (solar diurnal variation), 27-day cycle (solar rotation), 182.5-day cycle (Rieger-type periodicity in solar activity), and 1-year cycle (seasonal atmospheric changes) \cite{2021A&A...653A.146G,2016ApJ...817...38R}. The muon rate model for the simulation is expressed as:

\begin{equation*}
    R(t)= R(0)+ \sum_{i} a_{i} \sin{(\frac{2\pi t}{T_{i}}+\phi_{i})}
\end{equation*}

where \(R(t)\) represents the simulated muon rate, \(R(0)\) is the initial mock muon rate, $a_{i}$ is the amplitude, $\phi_{i}$ represents the phase, and  $T_{i}$ represents the periodicities of 1 day, 27 days, 182.5 days, and 1-year. To model the noise, we calculated the noise level as the standard deviation of the difference between the measured and smoothed muon rates:

\begin{equation*}
    \sigma_\text{noise}=\sqrt{ <(\text{R}_{\text{M15}}-\text{R}_{\text{M15,smooth}})^2>}
\end{equation*}

We used a one-dimensional Gaussian filter with a standard deviation ($\sigma$) of 100 for smoothing to obtain $\sigma_{\text{noise}}$. Fig.\,\ref{Fig_21} illustrates the results, where the blue line represents the actual variation of muon rates for M15 in 2011, and the red line shows the simulated data. The close match between these two lines demonstrates the accuracy of the simulation in replicating the behavior of the real data. Next, we take the ratio of M15 to the other 15 modules using the SG filter, as discussed in previous sections, to account for the detector effects in the remaining modules. The muon rate for the other 15 modules is given by: 

\begin{equation*}
    R_{i}(t)=\frac{R(t)}{r_{SG,i}}
\end{equation*}

\begin{figure}
    \centering
    \includegraphics[width=0.90\textwidth]{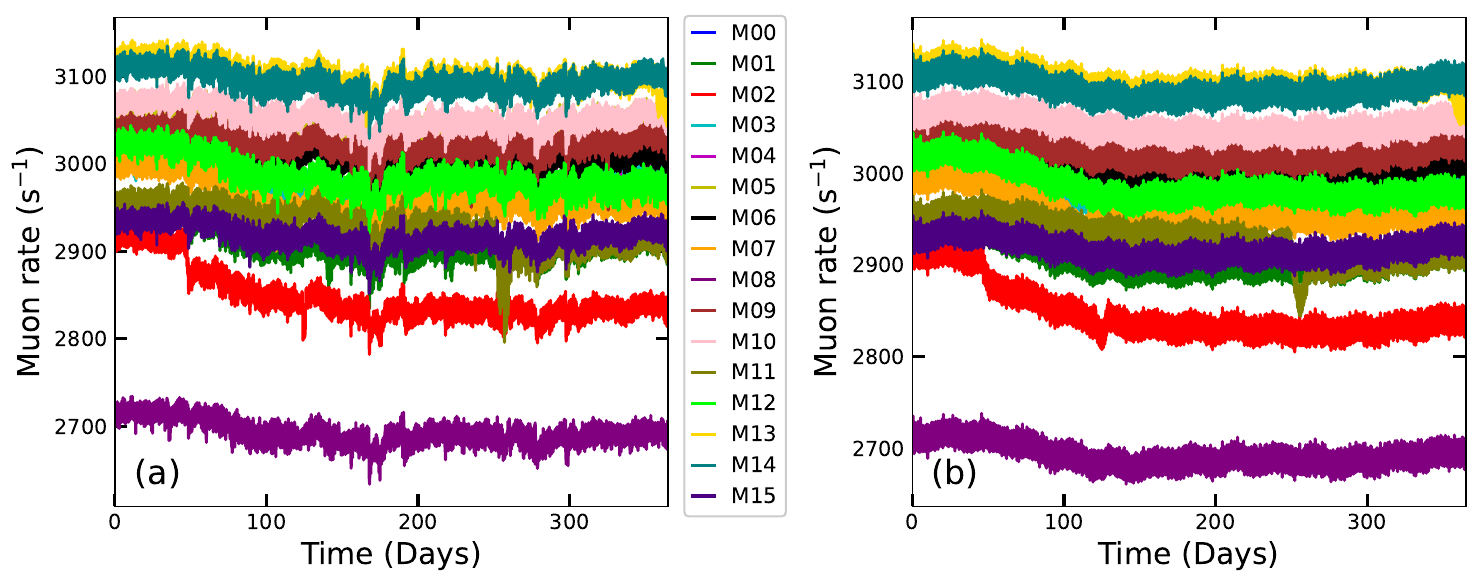}
    \caption{(a) Muon rate profiles for all 16 modules during 2011, and (b) the corresponding simulated data.}
    \label{Fig_22}
\end{figure}

$r_{SG,i}$ defines the ratio of M15 to i$^{th}$ module obtained using the SG filter, as discussed in the previous sections. Fig.\,\ref{Fig_22} shows the muon rate profiles for all 16 modules, where panel (a) presents the real muon rate data from 2011, and panel (b) shows the corresponding simulated data. The consistency between real and simulated data across all modules indicates that the simulation captures the relevant periodic behaviors and detector effects accurately.

\subsection{Comparison with legacy method}

\textbf{Legacy method:} M15 from 2011 is used as a reference for efficiency correction. In the simulation study, we use $R_{15}(t)$ as a static reference to correct the efficiency of the other modules, consistent with the legacy method \cite{Mohanty_2016_1,Mohanty_2017}.

\noindent \textbf{New method:} The new method dynamically selects the reference module by applying Bayesian block-based change point detection to refine the efficiency correction process. In the simulation, we calculate the efficiency corrections using a combination of $R_{i}(t)$ from the 16 modules, as outlined in earlier sections.

\begin{figure}
    \centering
    \includegraphics[width=0.90\textwidth]{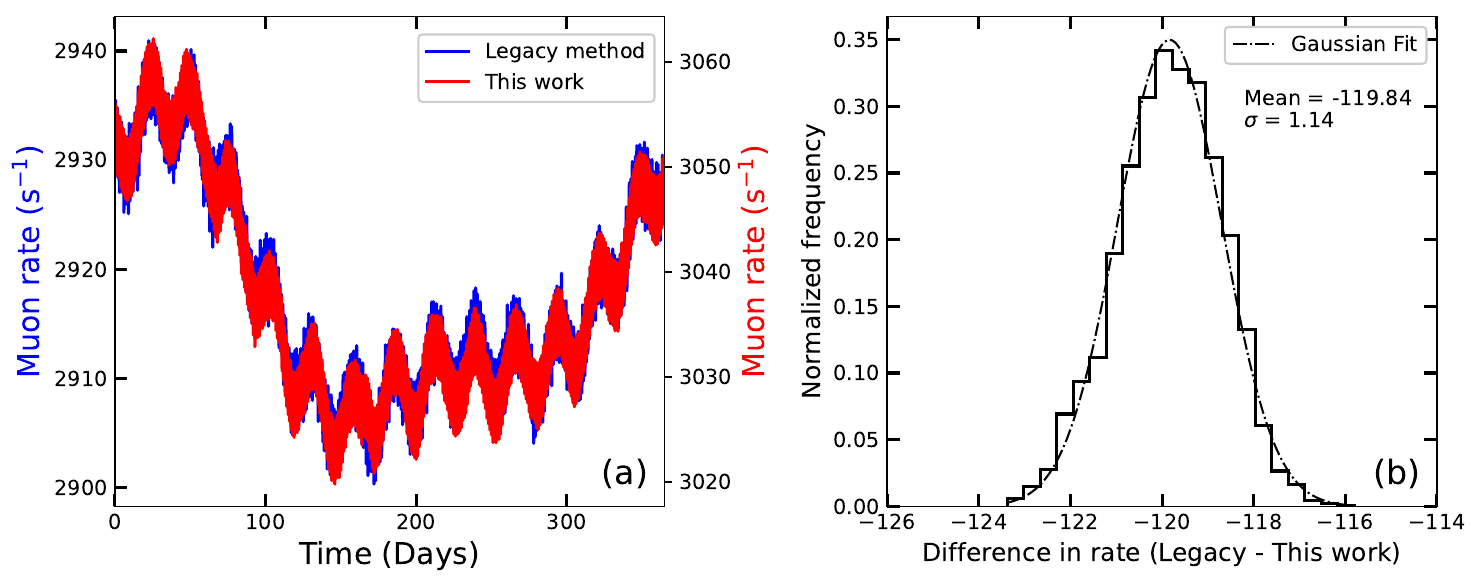}
    \caption{(a) The blue line represents the variation of efficiency-corrected muon rate using the legacy method, while the red line shows the same using the new method, both applied to the mock data. (b) Histogram of the rate differences between the two methods, fitted with a Gaussian distribution.}
    \label{Fig_23}
\end{figure}

Fig.\,\ref{Fig_23}(a) and (b) compare the two methods. In Fig.\,\ref{Fig_23}(a), the blue line shows the efficiency corrected muon rate using the legacy method, while the red line represents the same using the new method. The two methods exhibit strong agreement, with a correlation coefficient of 0.995, highlighting the improvement achieved with the new method, particularly in handling long-term efficiency variations. Fig.\,\ref{Fig_23}(b) shows the histogram of the difference in rates between the legacy method and this work.

This simulation-based analysis confirms the effectiveness of the new method, which dynamically adjusts efficiency corrections through Bayesian block detection, providing more accurate results compared to the old static reference approach. The strong correlation between the two methods validates the improvements and suggests that the new method is better suited for long-term CR studies where instrumental variations can affect data accuracy.

\subsection{Comparison with neutron monitor data}

In this section, the validation is based on the correlation between the muon rates from GRAPES-3 and the neutron rates from the neutron monitor database (NMDB), after correcting the muon data for atmospheric (pressure and temperature) and detector efficiency effects.  Since both muons and neutrons are produced in the upper atmosphere, their rates are influenced by atmospheric conditions, and by applying corrections, the residual variations in their rates should primarily reflect the space weather phenomena discussed in Sec.\,\ref{causes}. Therefore, a direct comparison of the muon rate from GRAPES-3 with the neutron rate from the neutron monitor provides an independent and meaningful validation framework for the new method against the legacy method.

The neutron monitor data used in this validation originates from the Princess Sirindhorn Neutron Monitor (PSNM) \cite{1964CaJPh..42.2443H}, located at Doi Inthanon, Thailand (18.59$^\circ$ N, 98.49$^\circ$ E), with a geomagnetic cutoff rigidity of 16.8 GV \cite{2016ApJ...817...38R}. This station was selected due to its comparable cutoff rigidity to GRAPES-3 ($\sim$17\,GV), ensuring a reliable reference for validation.

In both methods, the resulting muon rate is corrected for atmospheric pressure and efficiency, with only the atmospheric temperature correction left to be applied. To account for the effects of the upper atmospheric temperature, we apply equation \ref{eq:2}:

\begin{equation}\label{eq:2}
    \text{R}_{\text{T,P,}\epsilon}(\text{t}) = \frac{\text{R}_{\text{P,}\epsilon}(\text{t})}{(1+\alpha_\text{T} \Delta \text{T})}
\end{equation}

where $\text{R}_{\text{T,P,}\epsilon}(\text{t})$ represents the temperature, pressure, and efficiency-corrected muon rate, $\text{R}_{\text{P,}\epsilon}(\text{t})$ is the pressure and efficiency-corrected muon rate, and $\text{T}$ denotes the upper atmospheric temperature. Temperature data corresponding to the GRAPES-3 location is obtained from NASA’s MERRA-2 dataset \cite{NASA}, with the temperature coefficient set at $\alpha_\text{T}$ = --0.17 $\%\,\text{K}^{-1}$ \cite{Arunbabu_2017}. The pressure and efficiency-corrected neutron rate from the (PSNM) are publicly available via the NMDB \cite{PSNM}. As the upper atmospheric temperature has a minimal effect on neutron count variations, \cite{2016ApJ...817...38R}, no temperature correction is applied to the neutron data.

\begin{figure}[t]
    \centering
    \includegraphics[width=0.90\textwidth]{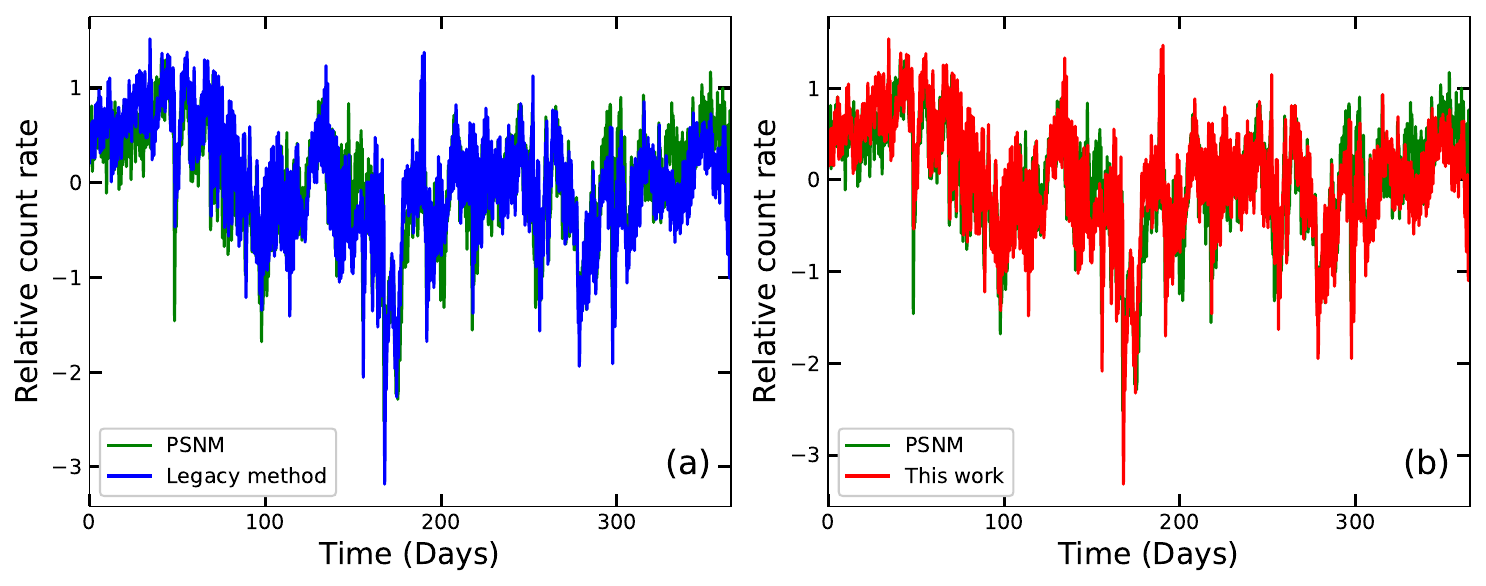}
    \caption{(a) The green line represents the relative neutron rate and the blue line represents the relative muon rate using the legacy method, after applying atmospheric and efficiency correction during 2011 and (b) the red line represents the relative muon rate using the new method. During 2011, both methods show a good correlation with the neutron data with correlation coefficents of 0.69 and 0.70 respectively.}
    \label{Fig_24}
\end{figure}

\begin{figure}[t]
    \centering
    \includegraphics[width=0.90\textwidth]{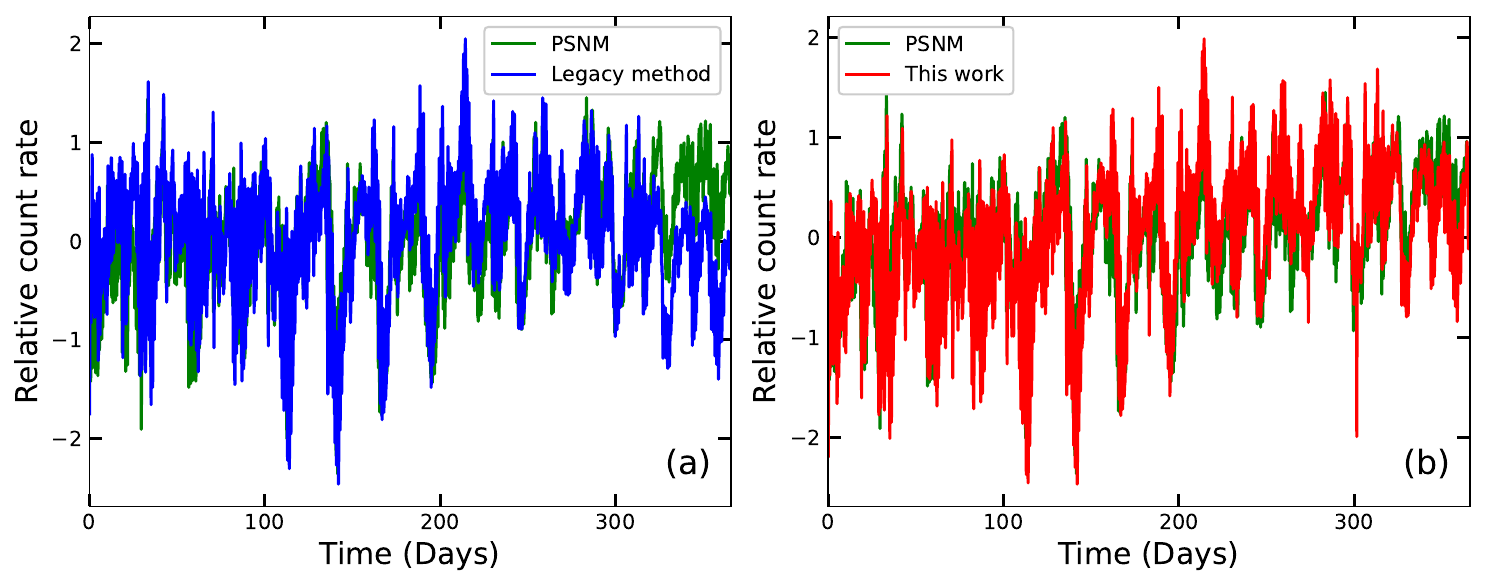}
    \caption{(a) The green line represents the relative neutron rate and the blue line represents the relative muon rate using the legacy method, after applying atmospheric and efficiency correction during 2016 and (b) the red line represents the relative muon rate using the new method. In 2016, the new method showed a good correlation with the neutron data compared to the legacy method, with correlation coefficients of 0.59 and 0.44, respectively.}
    \label{Fig_25}
\end{figure}

Fig.\,\ref{Fig_24}(a) shows the relative neutron rate (green) and the relative muon rate (blue) for 2011, determined using the legacy method. The atmospheric and efficiency corrections applied to both datasets resulted in a CC of 0.69. Fig.\,\ref{Fig_24}(b) shows the relative muon rate (red) using the new method, yielding a CC of 0.70. These results indicate that the two methods yield nearly identical results for 2011. Similarly, Fig.\,\ref{Fig_25}(a) and (b) depict the comparative analysis for 2016. The legacy method exhibits a correlation coefficient of 0.44, while the new method demonstrates a significant improvement with a CC of 0.59. This suggests that the new method is better at handling detector efficiency variations and extracting the underlying true muon flux.

The PSNM has been recording data since 2008, while the legacy method dataset spans 17 years, from 2001 to 2017. To ensure a fair comparison, we use data from 2008 to 2017. CCs are calculated for 10 years (2008-2017) to quantitatively assess the performance of both methods. Table \ref{tab:nm_correlation} presents the CCs between the muon rate observed from GRAPES-3 and the neutron rate from PSNM during 2008-2017. From 2008 to 2015, the two methods were consistent. However, significant improvements were observed in 2016 and 2017, where the correlation coefficients increased from 0.44 to 0.59 and from 0.29 to 0.48, respectively. These results indicate that the new method provides a better alignment with the independent neutron monitor data. This validation reinforces the robustness and adaptability of the new method over the legacy approach.

\begin{table}[h!]
\centering
\caption{Correlation coefficients between the muon rate observed from GRAPES-3 and neutron rate from PSNM.}
\begin{tabular}{|c|c|c|}
\hline
Year & Legacy & This work \\
\hline
2008 & 0.62 & 0.59 \\
2009 & 0.15 & 0.12 \\
2010 & 0.41 & 0.48 \\
2011 & 0.69 & 0.70 \\
2012 & 0.69 & 0.72 \\
2013 & 0.56 & 0.54 \\
2014 & 0.85 & 0.85 \\
2015 & 0.58 & 0.62 \\
2016 & 0.44 & 0.59 \\
2017 & 0.29 & 0.48 \\
\hline
\end{tabular}

\label{tab:nm_correlation}
\end{table}


\section{Summary and conclusions}

In this manuscript, we have presented a fully automated, algorithmic method to account for the long-term efficiency variations in the different modules of the GRAPES-3 muon telescope. The method employs Bayesian blocks to discretize the data into periods between change points, and examines the internal correlations between the 16 independent modules of G3MT to automatically identify the best 2 modules in each period, to then create a reference rate. This is in contrast to the legacy method, where the reference module was identified by the experimental operator based on their judgment. Unlike this legacy method which applied the efficiency corrections calendar year by calendar year, often introducing arbitrary discontinuities at the beginning of a calendar year, our new method dynamically selects the periods, ensuring continuity across calendar years. The method has been validated and compared against the legacy method on mock data, proving itself capable of matching the input time series more closely. Additionally, the corrected muon rates were compared with neutron monitor data, demonstrating a stronger correlation than the legacy method and further validating the reliability of the new approach.

The final corrected muon rate of all directions combined is shown in Fig.\,\ref{Fig_20}, spanning almost a complete solar magnetic cycle. This data is expected to be free from all the known instrumental effects, and carries valuable information of transient events as well as long-term CR variations due to various physical phenomena. This method can be separately applied for each of the 225 direction bins of G3MT. The method is suitable for any other experiment monitoring the long-term variation of naturally occurring particle flux, with sufficient redundancy in the number of detectors and can provide better signal-to-noise ratios in physics studies.


\section*{Acknowledgements}

We thank D.B. Arjunan, Late A.A. Basha, G.P. Francis, I.M. Haroon, V. Jeyakumar, Late S. Karthikeyan, S. Kingston,
N.K. Lokre, S. Murugapandian, S. Pandurangan, P.S. Rakshe, K. Ramadass, C. Ravindran, K.C. Ravindran, V. Santosh Kumar, S. Sathyaraj, M.S. Shareef, C. Shobana, R. Suresh Kumar, K. Viswanathan, and V. Viswanathan for their assistance over the years in running the experiment. 

\section*{Funding}

We acknowledge the support of the Department of Atomic Energy, Government of India, under Project Identification No. RTI4002.


\bibliography{07}
\end{document}